\documentclass[journal]{IEEEtran}
\usepackage{cite}
\usepackage{amsmath,amssymb,amsfonts}
\usepackage{algorithmic}
\usepackage{algorithm}
\usepackage{array}
\usepackage[caption=false,font=normalsize,labelfont=sf,textfont=sf]{subfig}
\usepackage{textcomp}
\usepackage{stfloats}   
\usepackage{graphicx}
\usepackage{bm}
\usepackage{cases}
\usepackage{xcolor}
\usepackage{url}
\usepackage{verbatim}
\newtheorem{theorem}{Theorem}[section]
\newtheorem{lemma}{Lemma}[section]
\usepackage{graphicx}
\usepackage{cite}
\usepackage{hyperref}
\begin{document}

\title{Constant-Modulus Secure Analog Beamforming for an IRS-Assisted Communication System with Large-Scale Antenna Array}

\author{Weijie Xiong, Jingran Lin, Zhiling Xiao, and Qiang Li
\thanks{This work was supported in part by the Natural Science Foundation of China (NSFC) under Grant 62171110. \textit{(Corresponding author: Jingran Lin.)}.}
\thanks{Jingran Lin and Qiang Li are with the School of Information and Communication Engineering, University of Electronic Science and Technology of China, Chengdu 611731, China, and also with the Laboratory of Electromagnetic Space Cognition and Intelligent Control, Beijing 100083, China, and the Tianfu Jiangxi Laboratory, Chengdu, Sichuan 641419, China (e-mail: jingranlin@uestc.edu.cn; lq@uestc.edu.cn).}
\thanks{Weijie Xiong and Zhiling Xiao are with the School of Information and Communication Engineering, University of Electronic Science and Technology of China, Chengdu 611731, China (e-mail: 202311012313@std.uestc.edu.cn; xiaozhiling@std.uestc.edu.cn).}
}

\markboth{Journal of \LaTeX\ Class Files,~Vol.~14, No.~8, August~2021}%
{Shell \MakeLowercase{\textit{et al.}}: A Sample Article Using IEEEtran.cls for IEEE Journals}


\maketitle

\begin{abstract}
Physical layer security (PLS) is an important technology in wireless communication systems to safeguard communication privacy and security between transmitters and legitimate users. The integration of large-scale antenna arrays (LSAA) and intelligent reflecting surfaces (IRS) has emerged as a promising approach to enhance PLS. However, LSAA requires a dedicated radio frequency (RF) chain for each antenna element, and IRS comprises hundreds of reflecting micro-antennas, leading to increased hardware costs and power consumption. To address this, cost-effective solutions like constant modulus analog beamforming (CMAB) have gained attention. This paper investigates PLS in IRS-assisted communication systems with a focus on jointly designing the CMAB at the transmitter and phase shifts at the IRS to maximize the secrecy rate. The resulting secrecy rate maximization (SRM) problem is non-convex. To solve the problem efficiently, we propose two algorithms: (1) the time-efficient Dinkelbach-BSUM algorithm, which reformulates the fractional problem into a series of quadratic programs using the Dinkelbach method and solves them via block successive upper-bound minimization (BSUM), and (2) the product manifold conjugate gradient descent (PMCGD) algorithm, which provides a better solution at the cost of slightly higher computational time by transforming the problem into an unconstrained optimization on a Riemannian product manifold and solving it using the conjugate gradient descent (CGD) algorithm. Simulation results validate the effectiveness of the proposed algorithms and highlight their distinct advantages.
\end{abstract}


\section{Introduction}
\IEEEPARstart{W}{ireless} communication networks are highly vulnerable to information leakage due to their open-access nature, making them susceptible to eavesdropping attacks \cite{khisti2010secure}. This challenge has driven the development of physical-layer security (PLS) techniques, originating from Wyner’s pioneering work in the 1970s \cite{6772207}. Wyner introduced the concept of exploiting spatial diversity to enhance security by directing beams toward legitimate receivers while creating interference for eavesdroppers. This enables reliable data decoding for legitimate users while making it difficult for eavesdroppers to extract meaningful information \cite{leung1978gaussian,yang2023secure,liu2024survey}. Since then, PLS has become a key research focus in wireless communication networks.

Despite advancements in PLS, emerging eavesdropping technologies, such as collaborative eavesdropping \cite{huang2021navigation,yuan2019secrecy}, UAV-assisted interception \cite{dai2022unmanned,adil2023uav}, and near-field eavesdropping \cite{zhang2024physical,chen2024physical}, disrupt the spatial diversity crucial for distinguishing legitimate users from eavesdroppers. Conventional methods, limited by the number of antennas at the base station, often fail to effectively address these threats \cite{bjornson2020scalable}. To overcome these limitations, large-scale antenna array (LSAA) systems, with significantly increased degrees of freedom (DoFs), offer a promising solution \cite{chen2020structured,tang2022dilated}. By enabling precise beamforming, LSAA systems restore spatial diversity and enhance spectral efficiency as well as communication security \cite{zhu2014secure,zhu2015linear,wu2016secure}. Specifically, the work in \cite{zhu2014secure} first analyzed PLS in a multicell LSAA-based system using maximum ratio transmission (MRT) beamforming and nullspace artificial noise (AN) under both perfect and imperfect channel state information (CSI). Building on this, \cite{zhu2015linear} proposed a more advanced beamforming design based on matrix polynomials. The study in \cite{wu2016secure} evaluated the impact of active eavesdropping in LSAA-based systems and derived secrecy rates using matched filter beamforming. Additionally, \cite{kapetanovic2015physical} proposed active eavesdropper detection schemes leveraging the large spatial degrees of freedom in LSAA systems. Despite these advantages, LSAA-based systems may face challenges in environments where eavesdropping channels are correlated with legitimate channels, potentially degrading communication security.

To address these challenges, intelligent reflecting surfaces (IRS) have gained significant attention in recent years \cite{wu2019towards}. An IRS is a software-controlled metasurface equipped with passive, digitally controlled reflecting elements that adjust the phase shift of impinging signals, creating a favorable propagation environment and offering additional optimization DoFs \cite{cheng2023ris,liu2021reconfigurable}. Specifically, integrating IRS into LSAA-based systems can suppress the correlation between legitimate and eavesdropping channels by dynamically adjusting phase shifts while enhancing signal quality through reflected signals. Motivated by this, many studies have explored enhancing PLS performance in IRS-assisted LSAA-based systems \cite{asaad2022secure,hong2020artificial,dong2020enhancing,yu2020robust,wang2022intelligent,cui2019secure}. For instance, \cite{asaad2022secure} investigates the secrecy rate achieved through joint beamforming design at the LSAA-based BS and phase shift optimization at the IRS. To further enhance performance, \cite{hong2020artificial} and \cite{dong2020enhancing} propose AN-aided joint transmission schemes for IRS-assisted LSAA-based systems, specifically targeting improved PLS in the presence of passive attacks. In more practical scenarios, the secrecy rate of systems with non-colluding and colluding eavesdroppers is maximized by jointly optimizing the beamforming and phase shift matrices, as proposed in \cite{yu2020robust} and \cite{wang2022intelligent}, respectively. Additionally, \cite{cui2019secure} examines the system with perfect CSI to mitigate passive attacks when the eavesdropping channels are stronger than the legitimate channels.

However, existing research on PLS in IRS-assisted LSAA-based systems predominantly relies on fully digital or hybrid implementations at the BS  \cite{ahmed2018survey,molisch2017hybrid,lin2019hybrid}, which require a substantial number of radio frequency (RF) chains. While hybrid approaches reduce RF chain requirements compared to fully digital methods, the demand still grows with the number of antennas, especially in LSAA systems \cite{sohrabi2016hybrid}. Additionally, the IRS typically consist of hundreds of reflecting elements made from specialized materials with specific electromagnetic properties, such as metamaterials or phase-change materials. These combined factors significantly increase hardware costs and power consumption, thereby constraining the practical applicability of such systems.

To address this issue, one potential solution is to employ beamforming networks based solely on simple analog components, such as phase shifters (PSs). These systems manipulate only the phase of each beamforming component while maintaining a constant magnitude, a technique known as constant modulus analog beamforming (CMAB). Due to its hardware efficiency, CMAB has been widely applied in IRS-assisted LSAA-based systems \cite{pradhan2020hybrid,xiu2021reconfigurable,di2020hybrid,li2022joint,zhou2021stochastic,chen2023irs}. For instance, \cite{pradhan2020hybrid} jointly optimized CMAB and IRS phase shifts to minimize the mean squared error between transmitted and received symbols in LSAA-based systems. In \cite{xiu2021reconfigurable}, power allocation, IRS phase shifts, and CMAB were jointly designed to maximize the sum rate in IRS-assisted LSAA systems. Additionally, studies such as \cite{di2020hybrid} and \cite{li2022joint} focused on sum rate maximization, while \cite{zhou2021stochastic} aimed to minimize the sum outage probability. Unlike most studies that focus on single-IRS systems, \cite{chen2023irs} investigated the joint design of CMAB and phase shifts in multi-IRS-assisted systems to reduce pilot overhead.

It is important to highlight that while CMAB has been widely used in IRS-assisted LSAA systems to reduce hardware costs and power consumption, its role in enhancing PLS remains underexplored. Given the importance of PLS in communication systems, this paper investigates its design in IRS-assisted LSAA systems utilizing CMAB. Specifically, we consider a scenario where a transmitter with LSAA transmits information to a multi-antenna legitimate receiver, while a multi-antenna eavesdropper attempts interception. Our objective is to jointly design the CMAB at the transmitter and the passive beamforming at the IRS to maximize the secrecy rate. The main contributions of this work are summarized as follows:
\begin{itemize}
\item Unlike previous studies on PLS in IRS-assisted LSAA-based systems that rely on fully digital or hybrid implementations at the BS, we focus on a transmitter architecture utilizing fully analog components. This approach can reduce both energy consumption and hardware costs associated with using a large number of RF chains.
\item To address the non-convex secrecy rate maximization problem, we first propose a time-efficient algorithm that combines the Dinkelbach method \cite{zappone2015energy} and the block successive upper-bound minimization (BSUM) method \cite{razaviyayn2013unified}. Specifically, the Dinkelbach method is first utilized to transform the fractional optimization problem into a sequence of quadratic programs, which are subsequently solved using the BSUM method by iteratively updating sub-problems with tight upper bounds.
\item We further propose the product manifold conjugate gradient descent (PMCGD) algorithm, which achieves a higher secrecy rate with only a slight increase in computational time. In this algorithm, we first transform the original problem into an unconstrained one on the Riemannian product manifold and then apply the conjugate gradient descent (CGD) algorithm to solve it.
\end{itemize}

The structure of this paper is as follows. Section II outlines the system model and the problem statement. In Section III, we introduce a time-efficient Dinkelbach-BSUM algorithm designed to address the problem, along with the development of a PMCGD algorithm that offers enhanced performance. Section IV presents the simulation results, and finally, Section V provides the conclusion of the paper.

The following notations are used throughout the paper. A vector and a matrix are represented by $\bf a$ and $\bf A$ respectively; $(\cdot)^T$, $(\cdot)^H$, and $(\cdot)^*$ represents the transpose, conjugate transpose, and conjugate of a matrix or vector, respectively; $\bf I$ refers to the identity matrix of suitable dimensions; ${\mathbb C}^N$ denotes the space of complex vectors of dimension $N$; the circularly symmetrix complex Gaussian distribution with mean $\mu $ and variance $\sigma^2 $ is denoted as $\mathcal{C N}(\mu, \sigma^2)$; $\text{Tr}({\bf A})$,  $||\cdot||_F$ and  $|\cdot|$ represents trace operator, Frobenius norm and absolute value; $\nabla f$ denotes the Euclidean gradient; ${\bf A}\succeq0$ denotes a positive semi-definite matrix; the phase of each element of a matrix is denoted as $\text{arg} ({\bf A})$; $\text{Real}(\cdot)$ denotes the real part of a complex matrix; ${\bf A}\odot{\bf B}$ represents Kronecker product.

\section{System model and problem formulation}

\begin{figure}
  \begin{center}
  \includegraphics[width=3in]{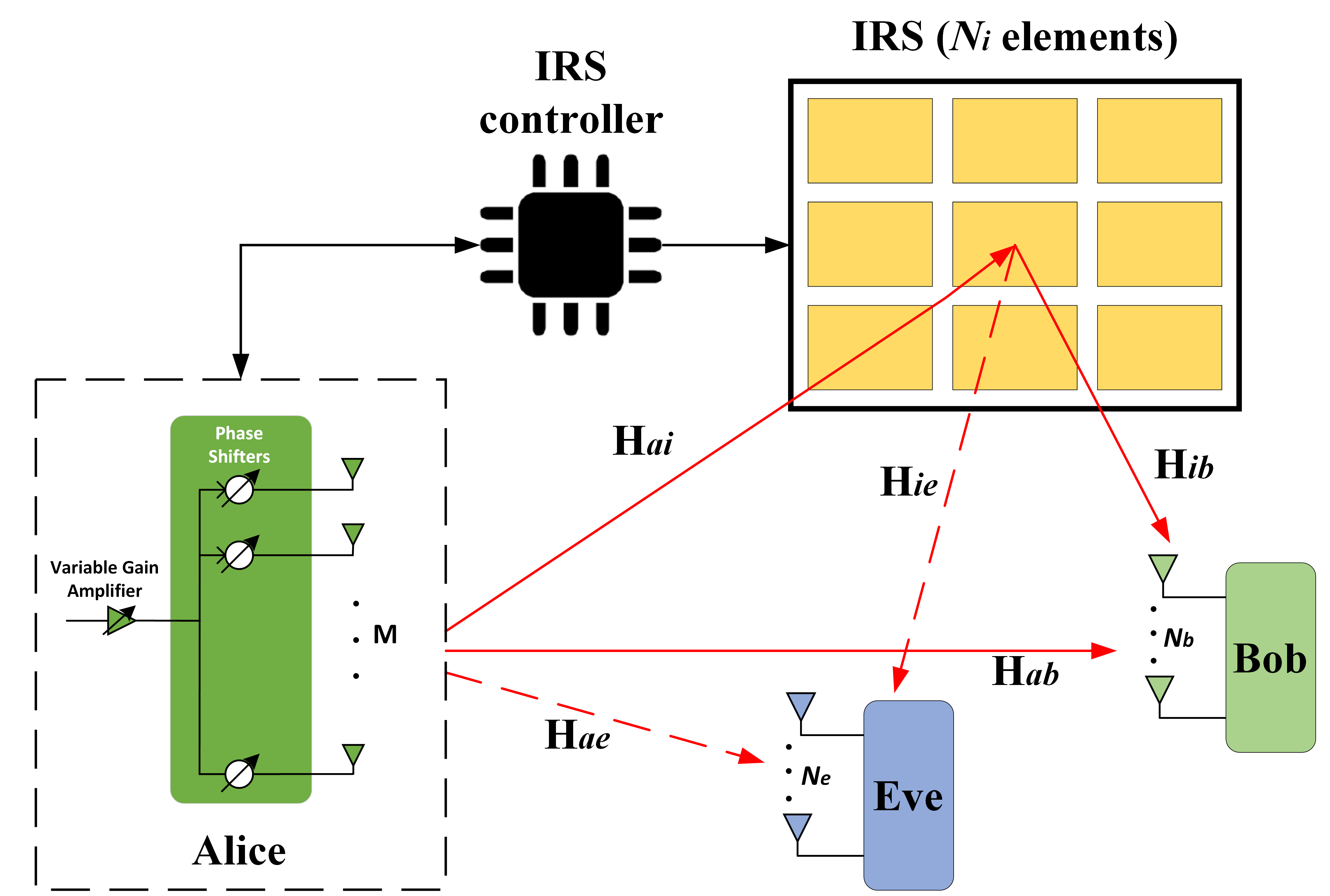}\\
  \caption{The IRS-assisted LSAA-based MIMO wiretap channel model.}\label{IRSLSAAS}
  \end{center}
\end{figure}

We consider an IRS-assisted LSAA-based MIMO wiretap channel model, as depicted in Fig. \ref{IRSLSAAS}. This model includes a transmitter (Alice), a receiver (Bob), an eavesdropper (Eve), and an IRS. The number of antennas deployed at Alice, Bob, and Eve is denoted by $M$, $N_b$, and $N_e$, respectively, while the IRS consists of $N_i$ reflecting elements. The transmitter Alice is equipped with a CMAB network driven by a common variable gain amplifier (VGA). Besides, the role of the IRS is to adjust the phase shift of its elements, passively directing signals from Alice to Bob and Eve. To analyze the performance limits of the IRS-assisted secrecy communication system under consideration, we assume that Alice and the IRS possess perfect knowledge of the global CSI for all relevant channels. This complete CSI enables them to jointly optimize the transmit and reflect beamforming strategies\footnote{It is reasonable to assume that \({\bf H}_{ae}\) and \({\bf H}_{ie}\) are known, particularly in scenarios where the eavesdropper acts as an active yet untrusted participant in the system \cite{xu2019resource,mukherjee2012detecting,li2019constant}. In such cases, the Eve may also function as a user of the system, with the transmitter designed to provide distinct services tailored to different user types, ensuring that these services are exclusively accessible to the intended users. Furthermore, for an active Eve, the CSI can be estimated from their transmissions. Interestingly, even in the case of a passive Eve, it may be possible to estimate the CSI by leveraging the local oscillator power inadvertently leaked from the receiver's RF frontend.}.

Let $s(t) \in \mathbb{C}$ represent the coded confidential information intended for Bob. Assuming $s(t)$ has unit power, the transmitted signal can be expressed as follows,
\begin{equation}
\mathbf{x}(t) = \mathbf{w}s(t),
\label{transmits}
\end{equation}
where $\mathbf{w} \in \mathbb{C}^{M}$ is the CMAB controlled by the PSs, i.e.,
\begin{equation}
|w_i| = P,\quad \forall i = 1,...,M,
\end{equation}
where $P > 0$ denotes the transmit power allocated per antenna, as established by the voltage gain amplifier (VGA).

Let $\mathbf{H}_{ab} \in \mathbb{C}^{N_b \times M}$, $\mathbf{H}_{ae} \in \mathbb{C}^{Ne \times M}$, $\mathbf{H}_{ai} \in \mathbb{C}^{N \times M}$, $\mathbf{H}_{ib} \in \mathbb{C}^{N_b \times N_i}$, and $\mathbf{H}_{ie} \in \mathbb{C}^{Ne \times N_i}$ represent the channel matrices for the direct links between Alice-Bob, Alice-Eve, Alice-IRS, IRS-Bob, and IRS-Eve, respectively. Let ${\bf{\Theta }} = diag({\bm \theta})$ denote the diagonal phase shift matrix for the IRS where ${\bm \theta} = [\theta_1,\theta_2,...,\theta_N]^T \in \mathbb{C}^{N_i}$. Assuming quasi-static channels, the received signals at Bob and Eve are expressed as,
\begin{subequations}
\begin{align}
& \mathbf{y}_b(t)=\mathbf{H}_{ab} \mathbf{x}(t) +\mathbf{H}_{ib} {\bf{\Theta }} \mathbf{H}_{ai} \mathbf{x}{(t)}+{\bf{n}}_b(t), \\
& \mathbf{y}_e(t)= {{\bf{H}}_{ae} {\bf{x}}(t) +{\bf{H}}_{ie} {\bf{\Theta }}  {\bf{H}}_{ai} \bf{x}}{(t)}+{\bf{n}}_e(t),
\end{align}
\label{reseivesigy}%
\end{subequations}
where ${\bf{n}}_b(t) \sim \mathcal{C N}(\mathbf{0}, \bf{I})$ and ${\bf{n}}_e(t) \sim \mathcal{C N}(\mathbf{0}, \bf{I})$ represent complex noise at Bob and Eve respectively.

Based on (\ref{transmits}) to (\ref{reseivesigy}), the transmission rates at Bob and Eve, denoted as $C_b(\bf{w},{\bm \theta})$ and $C_e(\bf{w},{\bm \theta})$, respectively, can be formulated as follows \cite{li2019constant},
\begin{subequations}
\begin{align}
& {C_b({\bf w},{\bm \theta})}= \log \left({1}+||({\mathbf H}_{ab} +{\mathbf H}_{ib} {\bf{\Theta }} {\mathbf H}_{ai}) {\bf w}||_2^2\right), \\
&  {C_e({\bf w},{\bm \theta})}= \log \left({1}+||({\mathbf H}_{ae} +{\mathbf H}_{ie} {\bf{\Theta }} {\mathbf H}_{ai}) {\bf w}||_2^2\right).
\end{align}
\label{trasnmittr}
\end{subequations}
Based on (\ref{trasnmittr}), the secrecy rate is formulated as \cite{li2019constant},
\begin{equation}
{C_s(\bf{w},{\bm \theta})}=\left[ {C_b(\bf{w},{\bm \theta})}- {C_e(\bf{w},{\bm \theta})} \right]^{+}, \label{UsetogCs}
\end{equation}
where $[u]^{+} \triangleq \max (u, 0)$ indicates the maximum number between 0 and $u$. The primary objective of this study is to determine the optimal CMAB $\bf w$ and the optimal phase shift ${\bm \theta}$ that maximize the secrecy rate $C_s(\bf{w},{\bm \theta})$. By simplifying the secrecy rate expression from equation (\ref{UsetogCs}) and omitting the log operation, the optimization problem can be reformulated as,
\begin{subequations}
\begin{align}
 \max _{{\bf w}, {\bm \theta}} &\quad \left\{  \frac{ {1}+||({\mathbf H}_{ab} +{\mathbf H}_{ib} {\bf{\Theta }} {\mathbf H}_{ai}) {\bf w}||_2^2}{{1}+||({\mathbf H}_{ae} +{\mathbf H}_{ie} {\bf{\Theta }} {\mathbf H}_{ai}) {\bf w}||_2^2},1 \right\}, \\
 \text { s.t. }&\quad\left|w_i\right|=P, \quad i=1, \ldots, M, \\
& \quad \left|\theta_j\right|=1, \quad j=1, \ldots, N_i .
\end{align}
\label{ttmaxinCs}%
\end{subequations}
By observing that the objective function in problem (\ref{ttmaxinCs}) yields the maximum between the constant 1 and the fraction $\frac{ {1}+||({\mathbf H}_{ab} +{\mathbf H}_{ib} {\bf{\Theta }} {\mathbf H}_{ai}) {\bf w}||_2^2}{{1}+||({\mathbf H}_{ae} +{\mathbf H}_{ie} {\bf{\Theta }} {\mathbf H}_{ai}) {\bf w}||_2^2}$, we note that only the fractional term depends on the variables $\bf w$ and $\bm \theta$. Therefore, it can be reformulated by converting the maximization to minimization and swapping the numerator and denominator in (\ref{ttmaxinCs}),
\begin{subequations}
\begin{align}
 \min _{\bf{w}, {\bm \theta}}& \quad   \frac{ {1}+||({\mathbf H}_{ae} +{\mathbf H}_{ie} {\bf{\Theta }} {\mathbf H}_{ai}) {\bf w}||_2^2}{{1}+||({\mathbf H}_{ab} +{\mathbf H}_{ib} {\bf{\Theta }} {\mathbf H}_{ai}) {\bf w}||_2^2},  \label{minttCsa}\\
 \text { s.t. }&\quad\left|w_i\right|=P, \quad i=1, \ldots, M, \label{useCs7b}\\
& \quad \left|\theta_j\right|=1, \quad j=1, \ldots, N_i .\label{useCs7c}
\end{align}
\label{minttCs}%
\end{subequations}
Solving problem (\ref{minttCs}) presents several challenges: firstly, the high dimensionality due to the use of massive antennas; secondly, the non-concavity of the objective function; thirdly, the non-convex nature of the constant modulus (CM) constraints on the $\bf{w}$ and ${\bm \theta}$ are highly coupled; and fourthly, the strong coupling between the variables $\bf{w}$ and ${\bm \theta}$.

Given these challenges, the subsequent sections will concentrate on developing practical methods to obtain high-quality approximate solutions for problem (\ref{minttCs}).

\section{Proposed Dinkelbach-BSUM Algorithm}
In this section, we propose a Dinkelbach-BSUM algorithm to solve the non-convex problem (\ref{minttCs}). Specifically, we first relax the fractional objective function (\ref{minttCsa}) into a sequence of quadratic programs using the Dinkelbach method. Subsequently, we provide a brief introduction to the BSUM method, followed by the proposal of a BSUM-based algorithm to address the converted problem.

\subsection{Problem Transformation by Dinkelbach method}
Problem (\ref{minttCs}) involves a ratio of quadratic functions with respect to $\bf w$ and $\bm \theta$, falling within the scope of fractional programming. Recent studies have extensively examined these types of problems in the context of energy efficiency applications, as detailed in \cite{zappone2015energy} and its references. The Dinkelbach method is a well-established approach for addressing fractional programming; for a comprehensive overview, readers are encouraged to consult \cite{zappone2015energy}. Here, we outline the key steps of the Dinkelbach method applied to problem (\ref{minttCs}), summarized in Algorithm 1. Essentially, this method transforms the fractional program (\ref{minttCs}) into a sequence of subproblems (\ref{NewminDM}) that are parameterized by $\alpha$.
\begin{algorithm}
	\floatname{algorithm}{Algorithm}
	\renewcommand{\algorithmicrequire}{\textbf{Input:}}
	\renewcommand{\algorithmicensure}{\textbf{Output:}}
	\caption{: The Dinkelbach method for the problem (\ref{minttCs}).}
	\label{algorithmD1}
	\begin{algorithmic}[1]
		\REQUIRE 
                $\bf w$, $\bm \theta$.\\
            \STATE {\textbf {Reapeat}}
            \STATE  \begin{equation} {\alpha} \leftarrow \frac{ 1+||({\mathbf H}_{ae} +{\mathbf H}_{ie} {\bf{\Theta }} {\mathbf H}_{ai}) {\bf w}||_2^2}{ 1+||({\mathbf H}_{ab} +{\mathbf H}_{ib} {\bf{\Theta }} {\mathbf H}_{ai}) {\bf w}||_2^2}, \notag\end{equation} 
            \STATE  \begin{subequations}
\begin{align}
 \min _{\bf{w}, {\bm \theta}}& \quad f({\bf w}, {\bm \theta}) =1+||({\mathbf H}_{ae} +{\mathbf H}_{ie} {\bf{\Theta }} {\mathbf H}_{ai}) {\bf w}||_2^2 \notag \\&\quad\quad\quad\quad\quad\quad-\alpha (1+||({\mathbf H}_{ab} +{\mathbf H}_{ib} {\bf{\Theta }} {\mathbf H}_{ai}) {\bf w}||_2^2),  \label{NewminDMa}\\
 \text { s.t. }&\quad\left|w_i\right|=P, \quad i=1, \ldots, M, \label{NewminDMb}\\
& \quad \left|\theta_j\right|=1, \quad j=1, \ldots, N_i . \label{NewminDMc}
\end{align}
\label{NewminDM}%
\end{subequations}
		\STATE {\textbf {Until }convergence}
            \ENSURE $\bf w$, $\bm \theta$.\\
	\end{algorithmic}%
\end{algorithm}

Although the Dinkelbach method simplifies the objective function by transforming it from fractional programming into a sequence of subproblems (\ref{NewminDM}), the task remains challenging due to the highly coupled variables $\bf w$ and $\bm \theta$ in (\ref{NewminDMa}), as well as the non-convex constraints in (\ref{NewminDMb}) and (\ref{NewminDMc}). We observe that when either $\bf w$ or $\bm \theta$ is fixed, the other forms a quadratic objective function, making it considerably easier to solve. Inspired by this, we exploit the partially quadratic nature of the objective function and propose an alternating minimization scheme based on the BSUM method. In the remainder of this section, we first briefly introduce the BSUM method, and then, based on this method, we propose a low-complexity solution to (\ref{NewminDM}).

\subsection{Brief introduction of the BSUM method}
Problem (\ref{NewminDM}) involves non-convex CM constraints (\ref{NewminDMb}) and (\ref{NewminDMc}) with respect to $\bf w$ and $\bm \theta$. These types of problems have been extensively studied in beamforming-related applications; see \cite{arora2021efficient} and references therein. The BSUM method is a classical approach commonly used to address such problems; for a comprehensive treatment, readers are referred to \cite{razaviyayn2013unified}. Here, we provide only a brief review of the BSUM method. When no ambiguity arises, certain notations are reused in this subsection. Consider the following problem,
\begin{equation}
\begin{aligned}
\min_{\bf w}& \quad f({\bf w}_1,{\bf w}_2,...,{\bf w}_n),\\
\text{s.t.}& \quad {{\bf w}_i} \in \chi_i,\; \forall i = 1,2,...,n,  
\end{aligned}
\end{equation}
where $f:\chi \to \mathbb{R}$ is the objective function may not be convex, $\chi=\chi_1 \times \chi_2 \times... \times\chi_n$ is the constraint set and ${\bf w}=({\bf w}_1,{\bf w}_2,...,{\bf w}_n)$ is organized into $n$ blocks, where each block ${{\bf w}_n}$ has dimensions $n_i\times1$. During the $k$-th iteration, the following sub-problem is addressed,
\begin{equation}
\begin{aligned}
&{\bf w}_i^{k+1} \in \arg \min_{{\bf w}_i} g_i ({\bf w}_i;{\bf w}_1^{k+1},...,{\bf w}_{i-1}^{k+1},{\bf w}_{i}^{k},...,{\bf w}_{n}^{k}),\\
&\text{s.t.} \quad {{\bf w}_i} \in \chi_i,\; \forall i = 1,2,...,n, 
\end{aligned}
\end{equation}
for all $i = 1, 2,...,n$, the blocks are updated in a cyclic manner, with ${\bf w}_i^k$ representing the update for block ${\bf w}_i$ during the $k$-th iteration. Analogous to the conventional majorization-minimization (MM) method, the function $g_i ({\bf w}_i;{\bf w}_1^{k+1},...,{\bf w}_{i-1}^{k+1},{\bf w}_{i}^{k},...,{\bf w}_{n}^{k})$ serves as a tight convex majorizer of the original objective function with respect to the block variable ${\bf w}_i$ and fulfills the following properties,
\begin{equation}
\begin{aligned}
& g_i ({\bf w}_i ; {\bf w}_i^k )  \geq f({\bf w}_i), \forall {\bf w}_i \in \chi_i, \\ 
& g_i ({\bf w}_i^k ; {\bf w}_i^k )  =f ({\bf w}_i^k ), \\ 
& \nabla g_i ({\bf w}_i^k ; {\bf w}_i^k )  =\nabla f({\bf w}_i^k) .
\end{aligned}
\label{Assumptiontub}
\end{equation}
Equation (\ref{Assumptiontub}) indicates that the surrogate function acts as a tight upper bound for the original objective function. This characteristic ensures that the objective function decreases with each iteration, ultimately converging to a stationary point of the original problem. Notably, the BSUM framework allows for the inclusion of nonconvex constraints \cite{arora2021efficient}. Additionally, the majorization applies only to objective functions for which the subproblems are challenging to minimize; otherwise, the original objective function is directly minimized for each block.

\subsection{Realization of the BSUM-based Algorithm}
By applying the BSUM method to problem (\ref{NewminDM}), the procedure results in two subproblems with respect to ${\bf w}$ and ${\bm \theta}$, as outlined below,
\begin{subequations}
\begin{align}
&{\bf w}^{k+1} = \arg \min_{{\bf w}_i=1,\forall i } \quad\quad\quad f({\bf w}, {\bm \theta}^k),\label{sunw1}\\
&{\bm \theta}^{k+1} = \arg \min_{{\bm \theta}_j=1,\forall j }  \quad\quad\quad f({\bf w}^{k+1}, {\bm \theta}).\label{suntheta1}
\end{align}
\end{subequations}
We now focus on developing an efficient algorithm for each sub-problem and subsequently updating them with a tight upper bound.

\subsubsection{Update $\bf w$} With $\bm \theta$ being fixed, the sub-problem (\ref{sunw1}) related to $\bf w$ after ignoring the constant term is formulated as,
\begin{equation}
\begin{aligned}
 \min _{\bf{w}} &\quad {{{\bf{w}}^{H}} {\bf A} {\bf{w}}}, \\
 \text { s.t. } &\quad\left|w_i\right|=P, \quad i=1, \ldots, M .\\
\end{aligned} 
\label{Updatewnewcf}
\end{equation}
where ${\bf A} = (\mathbf{H}_{ae} +\mathbf{H}_{ie} {\bf{\Theta }} \mathbf{H}_{ai})^H(\mathbf{H}_{ae} +\mathbf{H}_{ie} {\bf{\Theta }} \mathbf{H}_{ai})-\alpha(\mathbf{H}_{ab} +\mathbf{H}_{ib} {\bf{\Theta }} \mathbf{H}_{ai})^H(\mathbf{H}_{ab} +\mathbf{H}_{ib} {\bf{\Theta }} \mathbf{H}_{ai})$. It is well established that problem (\ref{Updatewnewcf}) has a closed-form optimal solution, specifically the principal generalized eigenvector of $\bf A$, when considering the total power constraint \cite{5605343}. However, the introduction of CM constraints complicates problem (\ref{Updatewnewcf}), transforming it into an NP-hard challenge \cite{li2019constant}. To address this issue, we propose utilizing the BSUM method to develop a tightly upper-bounded majorizing function that satisfies the properties outlined in (\ref{Assumptiontub}) for the objective of problem (\ref{Updatewnewcf}). The construction of this majorizing function is based on the following lemma from \cite{song2015optimization, hunter2004tutorial}.

\begin{lemma} \label{lemma31}
\textit{(Lemma 1 in \cite{song2015optimization})} The quadratic function of the form ${\bf w}^H {\bf T} {\bf w}$, where ${\bf T}$ is a Hermitian matrix, can tightly upper-bounded at the point ${\bf w}_k$, satisfying the properties outlined in (\ref{Assumptiontub}), by the expression ${\bf w}^H {\bf S} {\bf w} + 2 \text{Real}({\bf w}^H({\bf T} - {\bf S}){\bf w}_k) + {\bf w}_k^H({\bf S} - {\bf T}){\bf w}_k$, with ${\bf S}$ being a Hermitian matrix such that ${\bf S} \succeq {\bf T}$.
\end{lemma}

The proof of Lemma \ref{lemma31} can be established through a second-order Taylor expansion, in which the Hessian matrix ${\bf T}$ is replaced by another Hermitian matrix ${\bf S}$ satisfying ${\bf S} \succeq {\bf T}$. This lemma is also referred to as the quadratic upper bound principle for general twice-differentiable functions, as discussed in \cite{lange2016mm}.

To derive the solution for the variable $\bf w$, we utilize Lemma \ref{lemma31} to establish a tight upper bound for the objective function, expressed as follows,
\begin{equation}
\begin{aligned}
{{\bf w}^H {\bf A} {\bf w}}  \le & \lambda_{\text{max}}({\bf A}){\bf w}^H{\bf w} \\+ & 2 \text{Real}({\bf w}^H({\bf A}-\lambda_{\text{max}}({\bf A}){\bf I}){\bf w}^k) \\ + & ({\bf w}^k)^H (\lambda_{\text{max}}({\bf A}){\bf I}-{\bf A})({\bf w}^k),
\end{aligned}
\label{maxminnew}
\end{equation}
where $\bf I$ denotes an $M \times M$ identity matrix, and the function 
$\lambda_{\text{max}}({\bf A})$ indicates the maximum eigenvalue of matrix $\bf A$. The first term on the right side of inequality (\ref{maxminnew}) is constant due to the CM constraints, while the third term does not depend on $\bf w$. By disregarding the constant terms on the right-hand side of (\ref{maxminnew}), the majorized problem for solving problem (\ref{Updatewnewcf}) with respect to the variable $\bf w$ can be expressed as follows,
\begin{equation}
\begin{aligned}
 \min _{\bf{w}}&\quad \text{Real}({\bf w}^H({\bf A}-\lambda_{\text{max}}({\bf A}){\bf I}){\bf w}^k), \\
 \text { s.t. } &\quad\left|w_i\right|=P, \quad i=1, \ldots, M .\\
\end{aligned} 
\label{usetubw}
\end{equation}
 It can be demonstrated that problem (\ref{usetubw}) has a closed-form solution given by,
 \begin{equation}
 {\bf w} = P e^{j\arg((\lambda_{\text{max}}({\bf A}){\bf I}-{\bf A}){\bf w}^k)}.\label{Closedsw}
 \end{equation}

\subsubsection{Update $\bm \theta$} With $\bf w$ being fixed, the sub-problem (\ref{suntheta1}) related to $\bm \theta$ after ignoring the constant term is equivalently expressed as, 
\begin{equation}
\begin{aligned}
\min _{{\bm \theta}} & \quad\text{Tr}\{{\bf{\Theta }}^H {\bf B} {\bf{\Theta }} {\bf C} + {\bf{\Theta }}^H {\bf D}^H + {\bf{\Theta }} {\bf D} \} \\ & \quad\quad\quad\quad\quad\quad\quad- \alpha\text{Tr}\{{\bf{\Theta }}^H {\bf E} {\bf{\Theta }} {\bf C} + {\bf{\Theta }}^H {\bf F}^H + {\bf{\Theta }} {\bf F} \},  \\  
 \text{s.t.}&\quad \left|\theta_j\right|=1, \quad j=1, \ldots, N_i .
\end{aligned}
\label{optimiznewtheta}%
\end{equation}
where,
\begin{subequations}
\begin{align}
{\bf B} &= {\bf H}_{ie}^H{\bf H}_{ie}, \\
{\bf C} &= {\bf H}_{ai}{\bf w}{\bf w}^H{\bf H}_{ai}^H, \\
{\bf D}&={\bf H}_{ai}{\bf w}{\bf w}^H{\bf H}_{ae}^H{\bf H}_{ie}, \\
{\bf E} &= {\bf H}_{ib}^H{\bf H}_{ib}, \\
{\bf F}&={\bf H}_{ai}{\bf w}{\bf w}^H{\bf H}_{ab}^H{\bf H}_{ib}.
\end{align}
\label{usesimplyt1}%
\end{subequations}
The objective function in (\ref{optimiznewtheta}) contains matrix trace terms that can be compactly expressed using the properties of the trace operator, given as \cite{zhong2023joint},
\begin{subequations}
\begin{align}
 &\text{Tr}\{{\bf{\Theta }}^H {\bf B} {\bf{\Theta }} {\bf C}\} = {\bm \theta}^H({\bf B} \odot {\bf C}^T){\bm \theta},\\
 &\text{Tr}\{{\bf{\Theta }}^H {\bf E} {\bf{\Theta }} {\bf C}\} = {\bm \theta}^H({\bf E} \odot {\bf C}^T){\bm \theta}
\end{align}
\label{usesimplyt2}%
\end{subequations}
Furthermore, by assuming ${\bf d} = {diag}({\bf D})$ and ${\bf f} = {diag}({\bf F})$, the terms involving ${\bf D}^H$ and ${\bf F}^H$ in the objective function of (\ref{optimiznewtheta}) can be reformulated in terms of the vector $\bm \theta$, i.e.,
\begin{subequations}
\begin{align}
 &\text{Tr}\{{\bf{\Theta }}^H {\bf D}^H\} = \text{Tr}\{{\bf{\Theta }} {\bf D}\}=\text{Real}({\bm \theta}^H{\bf d}^*),\\
 &\text{Tr}\{{\bf{\Theta }}^H {\bf F}^H\} = \text{Tr}\{{\bf{\Theta }} {\bf F}\}=\text{Real}({\bm \theta}^H{\bf f}^*).
\end{align}
\label{usesimplyt3}%
\end{subequations}
Based on (\ref{usesimplyt1}) to (\ref{usesimplyt3}), (\ref{optimiznewtheta}) can be further simplified as,
\begin{equation}
\begin{aligned}
\min _{{\bm \theta}} &\quad {\bm{\theta }}^H  (({\bf B}- \alpha {\bf E})  \odot {\bf C}^T ) {\bm{\theta }} + 2\text{Real}({\bm \theta}^H({\bf d}^*-\alpha {\bf f}^*)), \\  
 \text{s.t.} &\quad \left|\theta_j\right|=1, \quad j=1, \ldots, N_i .
\end{aligned}
\label{newmin19} %
\end{equation}
The first term of the problem (\ref{newmin19}) is also a quadratic function, which can obtain a tight upper bound by using the \textit{Lemma III.1} similar to (\ref{Updatewnewcf}). Thus, the majorized problem to solve problem (\ref{newmin19}) for variable $\bm \theta$ is formulated as,
\begin{equation}
\begin{aligned}
\min _{\bm{\theta}}  &\quad \text{Real}({\bm \theta}^H(({\bf P}-\lambda_{\text{max}}({\bf P}){\bf I}){\bm \theta}^k+({\bf d}^*-\alpha {\bf f}^*))), \\
 \text{s.t.} &\quad \left|\theta_j\right|=1, \quad j=1, \ldots, N_i,
\end{aligned} 
\label{Realmintheta}
\end{equation}
where ${\bf P}=({\bf B}- \alpha {\bf E})  \odot {\bf C}^T$. It can be shown that problem (\ref{Realmintheta}) admits the following closed-form solution,
\begin{equation}
{\bm \theta} = e^{j\arg((\lambda_{\text{max}}({\bf P}){\bf I}-{\bf P}){\bm \theta}^k-({\bf d}^*-\alpha {\bf f}^*))}.
\end{equation}

\subsection{Overview of the Dinkelbach-BSUM Algorithm}
The complete BSUM algorithm, detailing the steps, is presented in Algorithm \ref{algorithmBSUM2}. It is both simple and efficient, as each step can be computed analytically. By iteratively applying Algorithm \ref{algorithmD1} and Algorithm \ref{algorithmBSUM2}, a solution to problem (\ref{ttmaxinCs}) is ultimately achieved.

\begin{algorithm} 
	\floatname{algorithm}{Algorithm}
	\renewcommand{\algorithmicrequire}{\textbf{Input:}}
	\renewcommand{\algorithmicensure}{\textbf{Output:}}
	\caption{: The BSUM-based algorithm to the problem (8).}
	\label{algorithmBSUM2}
	\begin{algorithmic}[1]
		\REQUIRE 
                $\alpha$, Initialize with a feasible point $({\bf w}^0,{\bm \theta}^0)$, stopping criterion $\xi$ and set $k=0$.\\
            \STATE {\textbf {Reapeat}}
            \STATE  Calculate ${\bf w}^{k+1}$ by (16);
            \STATE  Calculate ${\bm \theta}^{k+1}$ by (21);
            \STATE $k \leftarrow k+1$;
		\STATE {\textbf {Until } stopping criterion (e.g. $|f({\bf w}^{k+1},{\bm \theta}^{k+1})-f({\bf w}^{k},{\bm \theta}^{k})| <\xi $) is satisfied.}\\
	\end{algorithmic}%
\end{algorithm}

\subsubsection{Analysis of computation complexity} The computation complexity of the Dinkelbach-BUSM algorithm is mainly contributed by the calculation of $\bf A$, $\lambda_{\text{max}}({\bf A})$, $\bf P$, $\lambda_{\text{max}}({\bf P})$, $\bf d$ and $\bf f$. The update of $\bf A$ and $\bf P$ both have complexity of order ${\cal O}((N_e+N_b)(M+N_i)N_i)$. The update of $\lambda_{\text{max}}({\bf A})$ has complexity of order ${\cal O}(M^3)$ and the update of $\lambda_{\text{max}}({\bf P})$ has complexity of order ${\cal O}(N_i^3)$. The update of $\bf d$ has the complexity of order ${\cal O}(N_eMN_i)$ and the update of $\bf f$ has the complexity of order ${\cal O}(N_bMN_i)$. Thus, the overall computation complexity is mainly about ${\cal O}(N_{T2}N_{T1}((N_e+N_b)(2MN+N_i^2)+M^3+N_i^3))$, where $N_{T1}$ and $N_{T2}$ is the number of iterations for the BUSM-based algorithm and the Dinkelbach method respectively.

\subsubsection{Analysis of convergence} To establish the convergence for the Dinkelbach-BUSM algorithm, we first prove the convergence of the BSUM-based algorithm to a stationary point in Theorem \ref{theorem1}.

\begin{theorem} \label{theorem1}
Let $\{{\bf w},{\bm \theta}\}$ be a sequence generated by Algorithm \ref{algorithmBSUM2}. Then, every limit point generated by this sequence is a stationary point of problem (\ref{NewminDM}).
\end{theorem}

\textit{Proof}: See Appendix \ref{appendixA}. $\hfill\blacksquare$

We are now ready to prove the convergence of the Dinkelbach-BSUM algorithm in Algorithm \ref{algorithmD1}. At each iteration, the BSUM-based algorithm ensures that the surrogate function \(f(\mathbf{w}, \boldsymbol{\theta})\) decreases (or remains constant). Consequently, the ratio between the numerator and denominator in the update of \(\alpha\) (both dependent on the optimized values of \(\mathbf{w}\) and \(\boldsymbol{\theta}\)) in step 2 of Algorithm \ref{algorithmD1} will not increase. Since \(\alpha\) is non-increasing, it follows that the objective function in problem (\ref{minttCs}) is also non-increasing. Furthermore, as the objective function is bounded below by 1, we can conclude that every point generated by the Dinkelbach-BSUM algorithm converges to a limit point of problem (\ref{minttCs}).

\section{Proposed PMCGD Algorithm}
Although the proposed algorithm in the previous section converges within a relatively small number of iterations, its performance, in terms of optimizing the secrecy rate, is slightly compromised due to the relaxation of the fractional objective function. Specifically, Step 2 in Algorithm \ref{algorithmD1} reformulates the non-convex fractional objective function into a weighted difference function at each iteration. This reformulation may result in a loss of the global geometric structure of the original objective function, which can lead to suboptimal solutions that deviate from the global optimum. To achieve better secure performance, a more efficient PMCGD algorithm without relaxation is proposed to solve the problem (\ref{minttCs}) by slightly sacrificing the convergence speed. Specifically, by interpreting the constraint spaces of $\bf w$ in (\ref{useCs7b}) and ${\bm \theta}$ in (\ref{useCs7c}) as complex circle manifolds, we establish a Riemannian product manifold. This allows the constrained optimization problem (\ref{minttCs}) to be reformulated as an unconstrained optimization problem over this product manifold. Additionally, the CGD algorithm can be extended to operate on manifolds following specific guidelines. The details of the algorithm's development are outlined below.

\subsection{Construction of Product Manifold}

Generally, the CM constraints (\ref{useCs7b}) and (\ref{useCs7c}) in the Euclidean space can be interpreted as restricted search regions in the Riemannian space, denoted as \(\mathcal{M}_{\mathbf{w}}\) and \(\mathcal{M}_{\boldsymbol{\theta}}\), respectively. These regions are collectively referred to as complex circle manifolds, which have been extensively studied in applications related to waveform design; see \cite{yu2016alternating,hu2022constant} and references therein.

As described in \cite{boumal2023introduction}, multiple individual manifolds can be combined into a single manifold, known as a product manifold. Specifically, the two manifolds \(\mathcal{M}_{\mathbf{w}}\) and \(\mathcal{M}_{\boldsymbol{\theta}}\) can be combined to form a product manifold \(\mathcal{M}\), defined as,
\begin{equation}
\mathcal{M} = \mathcal{M}_{\mathbf{w}} \times \mathcal{M}_{\boldsymbol{\theta}} = \left\{(\mathbf{w}, \boldsymbol{\theta}) : \mathbf{w} \in \mathcal{M}_{\mathbf{w}}, \boldsymbol{\theta} \in \mathcal{M}_{\boldsymbol{\theta}} \right\},
\end{equation}
where \(\mathcal{M}\) is the product manifold with dimension \(M + N_i\). The individual complex circle manifolds \(\mathcal{M}_{\mathbf{w}}\) and \(\mathcal{M}_{\boldsymbol{\theta}}\) are defined as,
\begin{subequations}
\begin{align} 
& \mathcal{M}_{\mathbf{w}} = \left\{\mathbf{w} \in \mathbb{C}^M \mid |w_i| = P, \quad i = 1, \ldots, M \right\}, \\ 
& \mathcal{M}_{\boldsymbol{\theta}} = \left\{\boldsymbol{\theta} \in \mathbb{C}^{N_i} \mid |\theta_j| = 1, \quad j = 1, \ldots, N_i \right\}.
\end{align}
\end{subequations}
Generally, the product manifold \(\mathcal{M}\) combines the structural and constraint-related characteristics of \(\mathcal{M}_{\mathbf{w}}\) and \(\mathcal{M}_{\boldsymbol{\theta}}\), enabling joint optimization over both \(\mathbf{w}\) and \(\boldsymbol{\theta}\) while preserving the individual constraints. For a more detailed explanation of product manifolds and their applications, readers are referred to Chapter 3 of \cite{boumal2023introduction}, which provides a comprehensive discussion on the topic. This framework is a powerful tool for solving optimization problems involving multiple coupled variables constrained to distinct manifolds.

Then, the constrained optimization problem over the Euclidean space can transform into an unconstrained optimization problem over the Riemannian space, i.e.,
\begin{equation}
\min _{({\bf{w}}, {\bm \theta})\in \mathcal{M}} \quad f({\bf w},{\bm \theta})=\frac{ {1}+||({\mathbf H}_{ae} +{\mathbf H}_{ie} {\bf{\Theta }} {\mathbf H}_{ai}) {\bf w}||_2^2}{{1}+||({\mathbf H}_{ab} +{\mathbf H}_{ib} {\bf{\Theta }} {\mathbf H}_{ai}) {\bf w}||_2^2}. \label{maniCsmin}
\end{equation}

\subsection{Construction of Tangent Space}
The product manifold \(\mathcal{M}\) combines the properties of individual manifolds, enabling joint optimization while preserving the original constraints. However, due to its inherent curvature and nonlinearity, the product manifold \(\mathcal{M}\) presents significant challenges for algorithm design when working directly on the manifold. To address these challenges, constructing the tangent space of the product manifold becomes essential. The tangent space serves as a local linear approximation to the curved manifold, providing a Euclidean-like environment where traditional optimization algorithms can be adapted to navigate and optimize over the manifold's structure effectively.

Specifically, the tangent space \({\mathcal T}_{({\bf w},{\bm \theta})}{\mathcal M}\) of the product manifold \({\mathcal M}\) at a point \(({\bf w}, {\bm \theta})\) is generally expressed as \cite{boumal2023introduction},
\begin{equation}
\begin{aligned}
{\mathcal T}_{({\bf w},{\bm \theta})}{\mathcal M} &= \mathcal{T}_{\mathbf{w}} \mathcal{M}_{\bf w} \times \mathcal{T}_{\bm \theta} \mathcal{M}_{\bm \theta} \\
&= \left\{\begin{array}{l}
({\bm \xi}_{\mathbf{w}} \in \mathbb{C}^M, {\bm \xi}_{\bm{\theta}} \in \mathbb{C}^{N_i}): \\
\Re\{{\bm \xi}_{\mathbf{w}} \odot \mathbf{w}^*\} = \mathbf{0}_M, \\
\Re\{{\bm \xi}_{\bm{\theta}} \odot \bm{\theta}^*\} = \mathbf{0}_{N_i}
\end{array}\right\},
\end{aligned}
\label{usetangentspace}%
\end{equation}
where \({\bm \xi}_{\mathbf{w}}\) and \({\bm \xi}_{\bm{\theta}}\) are tangent vectors at the points \(\mathbf{w}\) and \(\bm{\theta}\), respectively. In this formulation, the tangent space \({\mathcal T}_{({\bf w},{\bm \theta})}{\mathcal M}\) provides a direct way to approximate the curved manifold locally, enabling the use of optimization techniques that rely on linear operations while maintaining the manifold's constraints. This facilitates efficient optimization over the product manifold while respecting its geometric structure.

\begin{figure}[htbp]
  \begin{center}
  \includegraphics[width=2.5in]{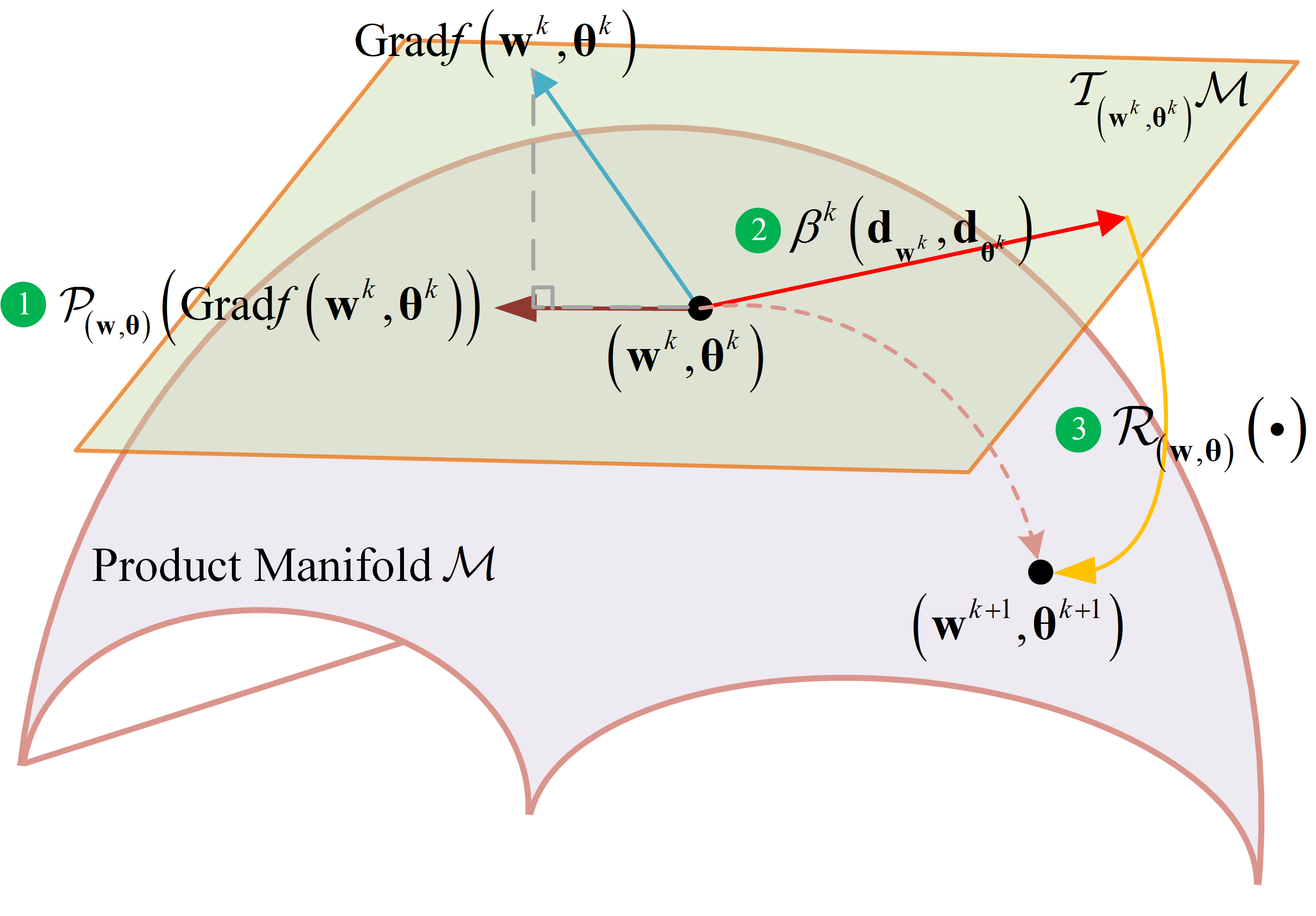}\\
  \caption{Visualization of gradient descent algorithm on manifolds.}\label{Vgdom}
  \end{center}
\end{figure}

From this point forward, the CGD algorithm is applied within the tangent space \({\mathcal T}_{({\bf w},{\bm \theta})}{\mathcal M}\) to solve problem (\ref{maniCsmin}). By leveraging the geometric properties of the product manifold, the CGD algorithm is specifically designed to optimize effectively within its tangent space. The procedure of the CGD algorithm at the \(k\)-th iteration is summarized in the following three steps, as illustrated in Fig. \ref{Vgdom}.

\textit{Step 1:} Calculate the corresponding Riemannian gradient $\text{grad}f({\bf w}^k,{\bm \theta}^k)$, and it can be obtained by projecting the Euclidean gradient $\text{Grad}f({\bf w}^k,{\bm \theta}^k)$ to the tangent space ${\mathcal T}_{({\bf w}^k,{\bm \theta}^k)}{\mathcal M}$ by the means of a projection operator ${\mathcal P}_{({\bf w},{\bm \theta})}(\cdot)$.

\textit{Step 2:} Apply conjugate gradient descent on the tangent space ${\mathcal T}_{({\bf w}^k,{\bm \theta}^k)}{\mathcal M}$ with descent direction $({\bf d}_{{\bf w}^k},{\bf d}_{{\bm \theta}^k})$ and step size $\beta^k$.

\textit{Step 3:} Obtain the feasible solution $({\bf w}^{k+1},{\bm \theta}^{k+1})$ by mapping the updated value from the tangent space back to the manifold ${\mathcal M}$, that is, ${\mathcal R}_{({\bf w},{\bm \theta})}(\cdot): {\mathcal T}_{({\bf w}^k,{\bm \theta}^k)}{\mathcal M} \to {\mathcal M}$. 

In the following, we give the details of the algorithm development at the $k$-th iteration based on the above three steps.

\subsection{CGD Algorithm over the Riemannian Manifold}

\subsubsection{Calculation of the Riemannian gradient}

Since the optimization is conducted within the tangent space \({\mathcal T}_{({\bf w}^k,{\bm \theta}^k)}{\mathcal M}\), a projection operation is required to map the Euclidean gradient \(\text{Grad}f({\bf w}^k, {\bm \theta}^k)\) onto the tangent space \({\mathcal T}_{({\bf w}^k, {\bm \theta}^k)}{\mathcal M}\). This projection ensures that the resulting gradient adheres to the geometric constraints of the manifold, yielding the Riemannian gradient \(\text{grad}f({\bf w}^k, {\bm \theta}^k)\). The Riemannian gradient is defined as follows \cite{boumal2023introduction},
\begin{equation}
\begin{aligned}
\text{grad}f({\bf w},{\bm \theta}) &= {\mathcal P}_{({\bf w},{\bm \theta})}(\text{Grad}f({\bf w},{\bm \theta}))\\
&=({\mathcal P}_{{\bf w}}(\text{Grad}_{{\bf w}}f({\bf w},{\bm \theta})),{\mathcal P}_{{\bm \theta}}(\text{Grad}_{{\bm \theta}}f({\bf w},{\bm \theta})))\\
&=(\text{grad}_{\bf w}f({\bf w},{\bm \theta}),\text{grad}_{\bm \theta}f({\bf w},{\bm \theta})),
\end{aligned}
\end{equation}
where \(\text{Grad}_{\bf w}f({\bf w},{\bm \theta})\) and \(\text{Grad}_{\bm \theta}f({\bf w},{\bm \theta})\) represent the corresponding Euclidean gradients. The projection operators \({\mathcal P}_{\bf w}(\cdot)\) and \({\mathcal P}_{\bm \theta}(\cdot)\) ensure that the Riemannian gradients lie entirely within their respective tangent spaces by removing components orthogonal to the manifolds. Furthermore, \(\text{grad}_{\bf w}f({\bf w},{\bm \theta})\) and \(\text{grad}_{\bm \theta}f({\bf w},{\bm \theta})\) denote the Riemannian gradients with respect to \({\bf w}\) and \({\bm \theta}\). Based on the tangent space defined in (\ref{usetangentspace}), they can be expressed as,
\begin{subequations}
\begin{align}
\text{grad}_{\bf w}&f({\bf w},{\bm \theta}) = {\mathcal P}_{{\bf w}}(\text{Grad}_{\bf w}f({\bf w},{\bm \theta}))\notag\\ &= \text{Grad}_{\bf w}f({\bf w},{\bm \theta})-\text{Real}\{\text{Grad}_{\bf w}f({\bf w},{\bm \theta}) \odot {\bf w}^* \} \odot {\bf w}, \\ 
\text{grad}_{\bm \theta}&f({\bf w},{\bm \theta}) = {\mathcal P}_{\bm \theta}(\text{Grad}_{\bm \theta}f({\bf w},{\bm \theta}))\notag\\ &= \text{Grad}_{\bm \theta}f({\bf w},{\bm \theta})-\text{Real}\{\text{Grad}_{\bm \theta}f({\bf w},{\bm \theta}) \odot {\bm \theta}^* \} \odot {\bm \theta},
\end{align}
\end{subequations}
where \(\text{Real}\{\text{Grad}_{\bf w}f({\bf w},{\bm \theta}) \odot {\bf w}^*\} \odot {\bf w}\) is the orthogonal component of the Euclidean gradient for \(\mathcal{M}_{\bf w}\), subtracted to ensure \(\text{grad}_{\bf w}f({\bf w},{\bm \theta})\) lies within the tangent space. Similarly, the projection \({\mathcal P}_{\bm \theta}\) removes the orthogonal component \(\text{Real}\{\text{Grad}_{\bm \theta}f({\bf w},{\bm \theta}) \odot {\bm \theta}^*\} \odot {\bm \theta}\), ensuring \(\text{grad}_{\bm \theta}f({\bf w},{\bm \theta})\) respects the tangent space of \(\mathcal{M}_{\bm \theta}\). These projections preserve the manifold structure and constraints, enabling effective optimization. The Euclidean gradients \(\text{Grad}_{\bf w}f({\bf w},{\bm \theta})\) and \(\text{Grad}_{\bm \theta}f({\bf w},{\bm \theta})\) are derived according to the Lemma \ref{lemma41}.

\begin{lemma} \label{lemma41}
The Euclidean gradients \(\text{Grad}_{\bf w}f({\bf w},{\bm \theta})\) and \(\text{Grad}_{\bm \theta}f({\bf w},{\bm \theta})\), with respect to \(\mathbf{w}\) and \(\boldsymbol{\theta}\), are provided in equations (\ref{Gradwnow}) and (\ref{Gradthetanow}), respectively.
\end{lemma}

\textit{Proof}: See Appendix \ref{Calgra}. $\hfill\blacksquare$

\subsubsection{Conjugate Gradient descent on the tangent space}
Before we apply conjugate gradient descent on the tangent space ${\mathcal T}_{({\bf w}^k,{\bm \theta}^k)}{\mathcal M}$, we first revisit some basic knowledge about the conjugate gradient descent over the Euclidean space. Recall that the standard conjugate gradient descent over the Euclidean space updates,
\begin{equation}
    {\bf w}^{k+1} = {\bf w}^{k} + \beta^{k}{\bf d}^{k},\label{usecgdm}
\end{equation}
where $k$ indicates the $k$-th iteration of the algorithm, $\beta^{k}>0$ denotes the step size and ${\bf d}^{k}$ is the descent direction. Generally, the step size $\beta^{k}$ can be calculated by the backtracking Armijo line-search method \cite{202312} and the search direction ${\bf d}^{k}$ is calculated by,
\begin{equation}
{\bf d}^{k} = -   \text{Grad}_{{\bf w}^{k}}f({\bf w}^{k})   +\sigma^{k} {\bf d}^{k-1}  
\end{equation}
where $\sigma^{k}$ is the conjugate parameter determined by the Polak–Ribiere rule \cite{babaie2015hybridization}, $\text{Grad}_{{\bf w}^{k}}f({\bf w}^{k})$ represents the Euclidean gradient of the parameter ${\bf w}^{k}$, and ${\bf d}^{k-1}$ is the descent direction from the previous iteration. Since the descent direction depends on both the current gradient and the previous descent direction, the algorithm ensures that the objective function non-increasing in each iteration. 

Typically, the difference between conjugate gradient descent over the tangent space ${\mathcal T}_{({\bf w}^k,{\bm \theta}^k)}{\mathcal M}$ and that over the Euclidean space lies in the updating of the descent direction ${\bf d}^{k}$. Specifically, since the two descent direction vectors ${\bf d}^{k-1}$ and ${\bf d}^{k}$ reside in two different tangent spaces, ${\mathcal T}_{({\bf w}^{k-1},{\bm \theta}^{k-1})}{\mathcal M}$ and ${\mathcal T}_{({\bf w}^{k},{\bm \theta}^{k})}{\mathcal M}$, respectively, they cannot be combined directly. To maintain consistency between iterations on a manifold, a mapping operator, \(\text{Trans}_{({\bf w}^{k-1},{\bm \theta}^{k-1}) \to ({\bf w}^{k},{\bm \theta}^{k})}(\cdot)\), is introduced. This operator maps the previous descent direction from the tangent space \({\mathcal T}_{({\bf w}^{k-1},{\bm \theta}^{k-1})}{\mathcal M}\) at \(({\bf w}^{k-1}, {\bm \theta}^{k-1})\) to the tangent space \({\mathcal T}_{({\bf w}^{k},{\bm \theta}^{k})}{\mathcal M}\) at the current point \(({\bf w}^{k}, {\bm \theta}^{k})\), ensuring that the descent direction remains valid within the new tangent space. The transformation is computed as follows \cite{boumal2023introduction},
\begin{equation}
\begin{aligned}
\text{Trans}_{({\bf w}^{k-1},{\bm \theta}^{k-1}) \to ({\bf w}^{k},{\bm \theta}^{k})}({\bf d}_{{\bf w}^{k-1}},{\bf d}_{{\bm \theta}^{k-1}})=\\(\text{Trans}_{{\bf w}^{k-1} \to {\bf w}^{k}}({\bf d}_{{\bf w}^{k-1}}),\text{Trans}_{{\bm \theta}^{k-1} \to {\bm \theta}^{k}}({\bf d}_{{\bm \theta}^{k-1}})),
\end{aligned}
\end{equation}
where the components are defined as,
\begin{subequations}
\begin{align}
&\text{Trans}_{{\bf w}^{k-1} \to {\bf w}^{k}}({\bf d}_{{\bf w}^{k-1}}) = \notag\\&\quad\quad\quad\quad\quad\quad{\bf d}_{{\bf w}^{k-1}}-\text{Real}\{{\bf d}_{{\bf w}^{k-1}} \odot ({\bf w}^{k})^* \} \odot{\bf w}^{k}, \\
&\text{Trans}_{{\bm \theta}^{k-1} \to {\bm \theta}^{k}}({\bf d}_{{\bm \theta}^{k-1}}) = \notag\\&\quad\quad\quad\quad\quad\quad{\bf d}_{{\bm \theta}^{k-1}}-\text{Real}\{{\bf d}_{{\bm \theta}^{k-1}} \odot ({\bm \theta}^{k})^* \} \odot{\bm \theta}^{k}.
\end{align}
\end{subequations}
These transformations subtract the components of the previous descent directions, \({\bf d}_{{\bf w}^{k-1}}\) and \({\bf d}_{{\bm \theta}^{k-1}}\), that are orthogonal to the tangent space ${\mathcal T}_{({\bf w}^{k},{\bm \theta}^{k})}{\mathcal M}$ at \(({\bf w}^{k}, {\bm \theta}^{k})\), ensuring that the mapped directions lie entirely within the new tangent space. Based on this discussion, the updated search directions at the current point \(({\bf w}^{k}, {\bm \theta}^{k}) \in \mathcal{M}\) are formulated as follows,
\begin{subequations}
\begin{align}
{\bf d}_{{\bf w}^{k}} &= -\text{grad}_{\bf w}f({\bf w}^{k},{\bm \theta}^{k}) + \sigma^{k}\text{Trans}_{{\bf w}^{k-1} \to {\bf w}^{k}}({\bf d}_{{\bf w}^{k-1}}),\\
{\bf d}_{{\bm \theta}^{k}} &= -\text{grad}_{\bm \theta}f({\bf w}^{k},{\bm \theta}^{k}) + \sigma^{k}\text{Trans}_{{\bm \theta}^{k-1} \to {\bm \theta}^{k}}({\bf d}_{{\bm \theta}^{k-1}}).
\end{align}
\end{subequations}
where \(-\text{grad}_{\bf w}f({\bf w}^{k},{\bm \theta}^{k})\) and \(-\text{grad}_{\bm \theta}f({\bf w}^{k},{\bm \theta}^{k})\) represent the Riemannian gradients at the current point. The momentum terms, \(\sigma^{k}\text{Trans}_{{\bf w}^{k-1} \to {\bf w}^{k}}({\bf d}_{{\bf w}^{k-1}})\) and \(\sigma^{k}\text{Trans}_{{\bm \theta}^{k-1} \to {\bm \theta}^{k}}({\bf d}_{{\bm \theta}^{k-1}})\), incorporate the transformed previous descent directions. This transformation is crucial because the tangent spaces vary with the manifold's curvature, and without it, the momentum terms would not align with the new tangent spaces. By combining the Riemannian gradients and momentum terms, this approach ensures efficient optimization while maintaining consistency with the geometric structure of \(\mathcal{M}\).

\subsubsection{Update the feasible solution}
After the optimization on the tangent space, we are supposed to map the vectors from the tangent space onto the product manifold itself to obtain the feasible solution  $({\bf w}^{k+1},{\bm \theta}^{k+1})$. In this case, a retraction operator ${\mathcal R}_{({\bf w},{\bm \theta})}(\cdot)$ is introduced, and the feasible solution $({\bf w}^{k+1},{\bm \theta}^{k+1})$ is calculated as,
\begin{equation}
\begin{aligned}
({\bf w}^{k+1},{\bm \theta}^{k+1}) & = {\mathcal R}_{({\bf w},{\bm \theta})}({\bf w}^{k} +  \beta^{k}{\bf d}_{{\bf w}^{k}},{\bm \theta}^{k} + \beta^{k}{\bf d}_{{\bm \theta}^{k}})\\&=({\mathcal R}_{{\bf w}}( {\bf w}^{k} + \beta^{k}{\bf d}_{{\bf w}^{k}} ),{\mathcal R}_{{\bm \theta}}({\bm \theta}^{k} + \beta^{k}{\bf d}_{{\bm \theta}^{k}})),
\end{aligned}
\end{equation}
where ${\mathcal R}_{{\bf w}}(\cdot)$ and ${\mathcal R}_{{\bm \theta}}(\cdot)$ are the retraction operator on ${\mathcal M}_{\bf w}$ and ${\mathcal M}_{{\bm \theta}}$ respectively, denoted as,
\begin{subequations}
\begin{align}
{\mathcal R}_{{\bf w}}( {\bf w}^{k} + \beta^{k}{\bf d}_{{\bf w}^{k}} ) &= \frac{P[{\bf w}^{k} + \beta^{k}{\bf d}_{{\bf w}^{k}}]_i}{|[{\bf w}^{k} + \beta^{k}{\bf d}_{{\bf w}^{k}}]_i|},\\
{\mathcal R}_{\bm \theta}( {\bm \theta}^{k} + \beta^{k}{\bf d}_{{\bm \theta}^{k}} ) &= \frac{[{\bm \theta}^{k} + \beta^{k}{\bf d}_{{\bm \theta}^{k}}]_i}{|[{\bm \theta}^{k} + \beta^{k}{\bf d}_{{\bm \theta}^{k}}]_i|}.
\end{align}
\end{subequations}

\subsection{Overview of the PMCGD algorithm}
Based on the above discussion, the PMCGD algorithm for the problem (\ref{minttCs}) is summarized as Algorithm \ref{PMCGD}. Compared to Algorithm \ref{algorithmD1}, Algorithm \ref{PMCGD} achieves better performance by directly optimizing the original objective function and constraint space without relying on relaxation or reformulation. The PMCGD algorithm leverages gradient information combined with a conjugate direction update strategy to iteratively approach the optimal solution while preserving the global geometric structure of the objective function. Additionally, in its manifold-based extensions, the algorithm maps gradients to the tangent space, ensuring strict adherence to the original constraints and avoiding the information loss typically associated with relaxation. This approach enables PMCGD to retain the global characteristics of the objective function, facilitating the efficient resolution of the problem.

\begin{algorithm}
	\floatname{algorithm}{Algorithm}
	\renewcommand{\algorithmicrequire}{\textbf{Input:}}
	\renewcommand{\algorithmicensure}{\textbf{Output:}}
	\caption{: The PMCGD algorithm for the problem (7).}
	\label{PMCGD}
	\begin{algorithmic}[1]
		\REQUIRE 
                 Initialize with a feasible point $({\bf w}^0,{\bm \theta}^0)$, stopping criterion $\xi$ and set $k=0$.\\
            \STATE {\textbf {Reapeat}}
            \STATE  Calculate the Euclidean gradient $\text{Grad}f({\bf w}^k,{\bm \theta}^k)$ by (27);
            \STATE  Calculate the Riemannian gradient $\text{grad}f({\bf w}^k,{\bm \theta}^k)$ by (25);
            \STATE  Calculate the descent direction $({\bf d}_{{\bf w}^{k}},{\bf d}_{{\bm \theta}^{k}})$ by (33);
             \STATE Update next feasible point $({\bf w}^{k+1},{\bm \theta}^{k+1})$ by  (34);
             \STATE  $k \leftarrow k+1$;
		\STATE {\textbf {Until } stopping criterion (e.g. $|f({\bf w}^{k+1},{\bm \theta}^{k+1})-f({\bf w}^{k},{\bm \theta}^{k})| <\xi $)  satisfied.}\\
	\end{algorithmic}%
\end{algorithm}

\subsubsection{Analysis of computation complexity} The complexity of the PMCGD algorithm is mainly contributed by the calculation of the Euclidean gradient $\text{grad}_{\bf w}f({\bf w},{\bm \theta})$ and $\text{grad}_{\bm \theta}f({\bf w},{\bm \theta})$. The update of the $\text{grad}_{\bf w}f({\bf w},{\bm \theta})$ has complexity of order ${\cal O}((N_e+N_b)(M+N_i)N_i)$ and the update of the $\text{grad}_{\bm \theta}f({\bf w},{\bm \theta})$ has complexity of order ${\cal O}((N_e+N_b)(M^2+MN_i+N_i^2))$. Assuming that the iteration number is $N_{T3}$, the total computation complexity of the PMCGD algorithm is mainly about ${\cal O}(N_{T3}(N_e+N_b)(M^2+2MN_i+2N_i^2))$. In our simulations, we observed that while the computational complexity of each iteration of the PMCGD algorithm is slightly lower than that of the Dinkelbach-BSUM algorithm, the PMCGD requires more iterations to converge, resulting in a slower overall convergence speed, as demonstrated in the simulation results.

\subsubsection{Analysis of convergence} We prove the convergence of Algorithm \ref{PMCGD} to a stationary point in Theorem \ref{theorem2}.

\begin{theorem} \label{theorem2}
Let $\{{\bf w},{\bm \theta}\}$ be a sequence generated by Algorithm \ref{PMCGD}. Then, every limit point generated by this sequence is a stationary point of problem (\ref{maniCsmin}).
\end{theorem}

\textit{Proof}: See Appendix \ref{appendixB}. $\hfill\blacksquare$

\section{Numerical Results}
In this section, we provide simulation results to evaluate the performance of the secrecy rate maximized design for the proposed IRS-assisted LSAA-based system. For comparison, the proposed method is compared to the following benchmark schemes:
\begin{itemize}
\item Without IRS in \cite{li2019constant}: Secure beamforming design utilizing analog beamformers by optimizing \(\mathbf{w}\) without IRS.
\item Random IRS: Secure beamforming design utilizing analog beamformers in \cite{li2019constant}, with random IRS phase shifts.
\item Upper Bound IRS: Secure beamforming design utilizing digital beamformers with total power constraints for \(\mathbf{w}\) in \cite{cheng2023ris}, and selects \(\bm{\theta}\) calculated from SDR with a rank-one solution. The optimal values of \(\mathbf{w}\) and \(\bm{\theta}\) are obtained and used as an upper bound.
\item SDR IRS: Secure beamforming design utilizing analog beamformers and employs SDR technology \cite{5447068} to solve problem (\ref{minttCs}).
\end{itemize}

In the simulations, we use the following settings, similar to those in \cite{li2019constant}: the number of antennas at Alice is \(M = 64\), the number of antennas at Bob and Eve are \(N_b = N_e = 4\), the number of IRS reflecting elements is \(N_i = 100\), and the power at Alice is \(P = 0 \, \text{dB}\), unless otherwise specified. Following \cite{dong2020enhancing}, we analyze a fading environment where all channels have both large-scale and small-scale fading. The entries of the small-scale fading matrix are modeled as complex zero-mean Gaussian random variables with unit variance. The large-scale fading path loss is set at -30 dB for a reference distance of 1 m, with a path loss exponent of 3 for all links. The distances are specified as follows: Alice to Bob (80 m), Alice to IRS (30 m), Alice to Eve (80 m), IRS to Bob (40 m), and IRS to Eve (40 m). Simulation results are averaged over 1000 random fading realizations, conducted using MATLAB 2021 on a standard PC with a 3.8GHz Intel Core Ultra7 155H processor, 1TB SSD, and 32GB RAM.

\begin{figure}[htbp]
  \begin{center}
  \includegraphics[width=3.5in]{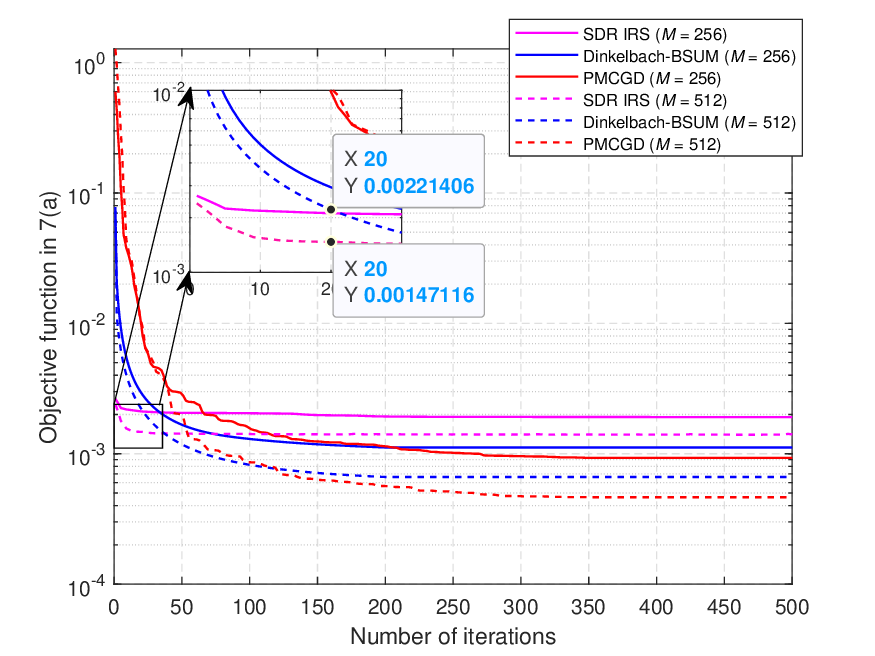}\\
  \caption{Convergence performance of the proposed methods.}\label{Convergcenoverall}
  \end{center}
\end{figure}

\begin{figure}[htbp]
  \begin{center}
  \includegraphics[width=3.5in]{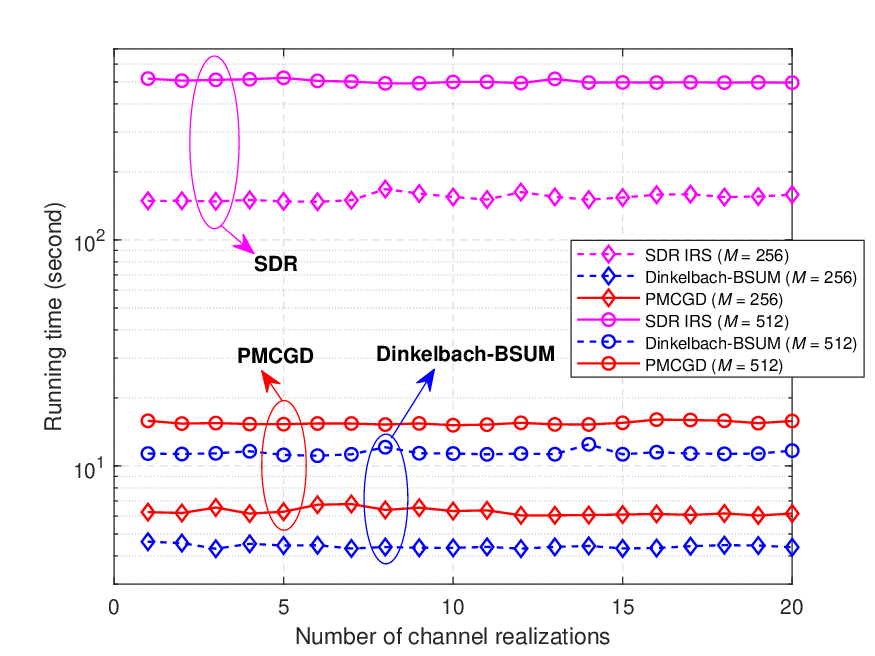}\\
  \caption{ Running time comparison between different methods for the first twenty randomly generated channels.}\label{Runningtime}
  \end{center}
\end{figure}

Fig. \ref{Convergcenoverall} illustrates the convergence performance of the Dinkelbach-BSUM, PMCGD, and SDR IRS algorithms, leading to several key observations. Firstly, the SDR IRS algorithm requires the fewest iterations to converge, around 20 iterations. This is followed by the Dinkelbach-BSUM algorithm, which converges in approximately 200 iterations, while the PMCGD algorithm takes the longest to converge, requiring around 350 iterations. It is important to note that, although the SDR IRS algorithm converges in the fewest iterations, it has significantly higher computational complexity per iteration, as it processes a large amount of data in each step. Consequently, this results in a much longer overall convergence time, as shown later in Fig. \ref{Runningtime}. Secondly, as the number of transmit antennas ($M$) increases, all algorithms achieve lower objective function values. This improvement arises from LSAA-based systems, where larger antenna arrays provide more DoFs for the system. Moreover, the increased number of antennas has little influence on the number of iterations required for convergence. Finally, the PMCGD algorithm attains a lower objective function value upon convergence compared to the Dinkelbach-BSUM algorithm for both $M=512$ and $M=256$. In contrast, the SDR IRS algorithm results in the highest objective function value, indicating the worst performance.

Fig. \ref{Runningtime} compares the running times for the first twenty randomly generated channels using the Dinkelbach-BSUM, PMCGD, and SDR IRS algorithms. As shown in the figure, the average running time for one randomly generated channel is 4.62 seconds for $M=256$ and 11.34 seconds for $M=512$ using the Dinkelbach-BSUM algorithm, making it the fastest in terms of overall convergence time. The PMCGD algorithm follows, requiring 6.24 seconds for $M=256$ and 15.84 seconds for $M=512$. Meanwhile, the SDR IRS algorithm shows significantly higher computation times of 149.01 seconds for $M=256$ and 517.62 seconds for $M=512$. Additionally, we also test the computation time per iteration for each algorithm. The Dinkelbach-BSUM algorithm takes 0.023 seconds per iteration for $M=256$ and 0.056 seconds for $M=512$. The PMCGD algorithm requires 0.018 seconds and 0.045 seconds per iteration for $M=256$ and $M=512$, respectively. In contrast, the SDR IRS algorithm takes much longer, with 7.45 seconds per iteration for $M=256$ and 25.88 seconds for $M=512$. The significant reduction in per-iteration computation time for the proposed algorithms, compared to the SDR IRS algorithm, is due to the closed-form solutions derived at each iteration and the distributed implementation capability.

\begin{figure}[htbp]
  \begin{center}
  \includegraphics[width=3.5in]{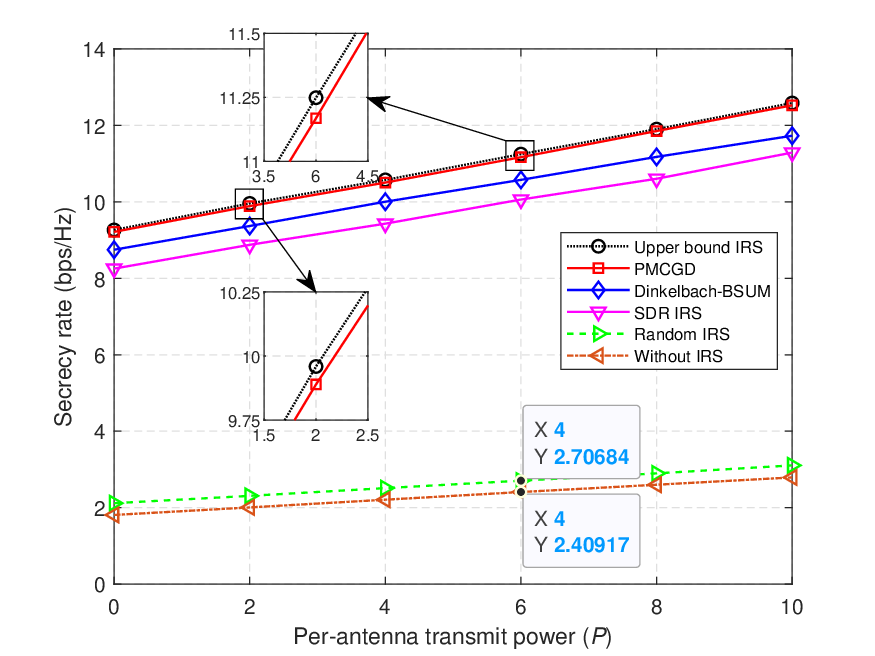}\\
  \caption{ Secrecy rate comparison among different strategies versus the per-antenna transmit power $P$.}\label{PapowerSR}
  \end{center}
\end{figure}

\begin{figure}[htbp]
  \begin{center}
  \includegraphics[width=3.5in]{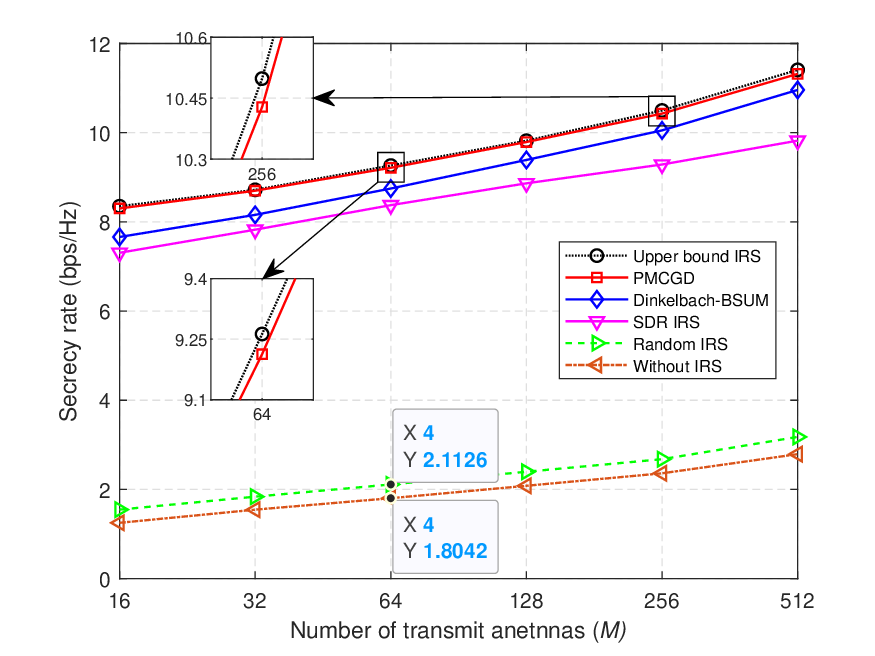}\\
  \caption{ Secrecy rate comparison among different strategies versus the number of transmit antennas at the base station $M$.}\label{TANCOM}
  \end{center}
\end{figure}

Figs. \ref{PapowerSR} and \ref{TANCOM} illustrate the average secrecy rate performance of various designs, with changes in per-antenna transmit power $P$ and the number of base station antennas $M$. As depicted in the figures, the average secrecy rate escalates with an increase in both Figs. \ref{PapowerSR} and \ref{TANCOM} for all methods. This improvement is attributable to the additional degrees of freedom and an expanded power limit provided to the system, underscoring the necessity of employing LSAA systems in scenarios with high propagation loss. Moreover, both the Dinkelbach-BSUM algorithm and the PMCGD algorithm outperform the SDR IRS algorithm, the Random IRS algorithm, and the Without IRS algorithm in terms of secrecy rate in both figures. This result shows that, compared to the SDR IRS algorithm, the proposed algorithms are better suited for LSAA-based systems with large data dimensions. Specifically, with $P=6$dB in Fig. \ref{PapowerSR}, the average secrecy rates for the PMCGD and Dinkelbach-BSUM algorithms are approximately 11.1 bps/Hz and 10.5 bps/Hz, respectively, exceeding those of the SDR IRS algorithm (10.0 bps/Hz), Random IRS algorithm (2.7 bps/Hz), and Without IRS algorithm (2.4 bps/Hz). Similarly, at $M=64$ in Fig. \ref{TANCOM}, the PMCGD and Dinkelbach-BSUM algorithms achieve average secrecy rates of about 9.2 bps/Hz and 8.7 bps/Hz, respectively, surpassing the SDR IRS algorithm (8.1 bps/Hz), Random IRS algorithm (2.1 bps/Hz), and Without IRS algorithm (1.8 bps/Hz). Additionally, the PMCGD algorithm approaches the efficiency of the Upper bound IRS algorithm regarding the secrecy rate.

\begin{figure}[htbp]
  \begin{center}
  \includegraphics[width=3.5in]{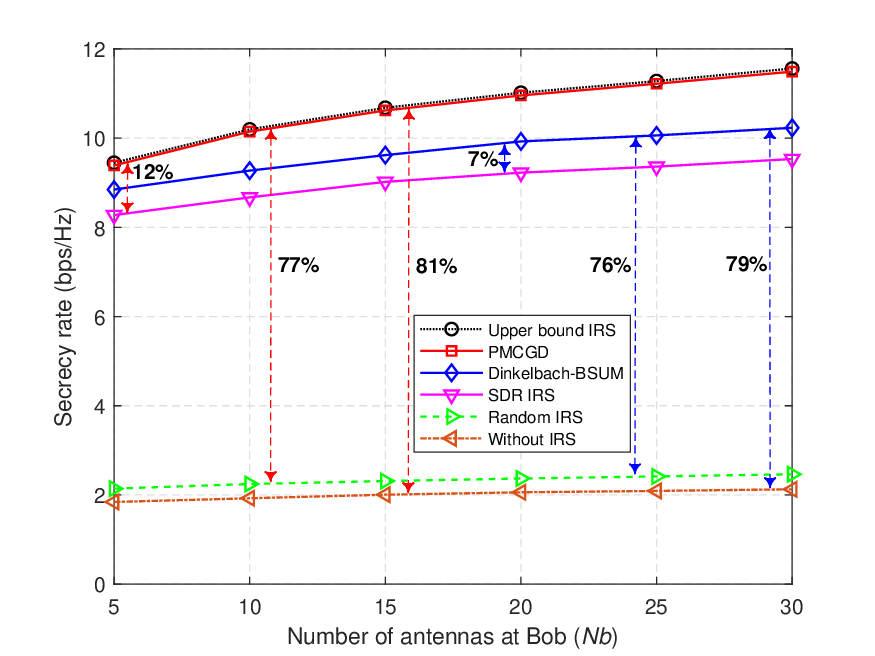}\\
  \caption{ Secrecy rate comparison among different strategies versus the number of antennas at Bob $N_b$.}\label{NumberatBOB}
  \end{center}
\end{figure}

\begin{figure}[htbp]
  \begin{center}
  \includegraphics[width=3.5in]{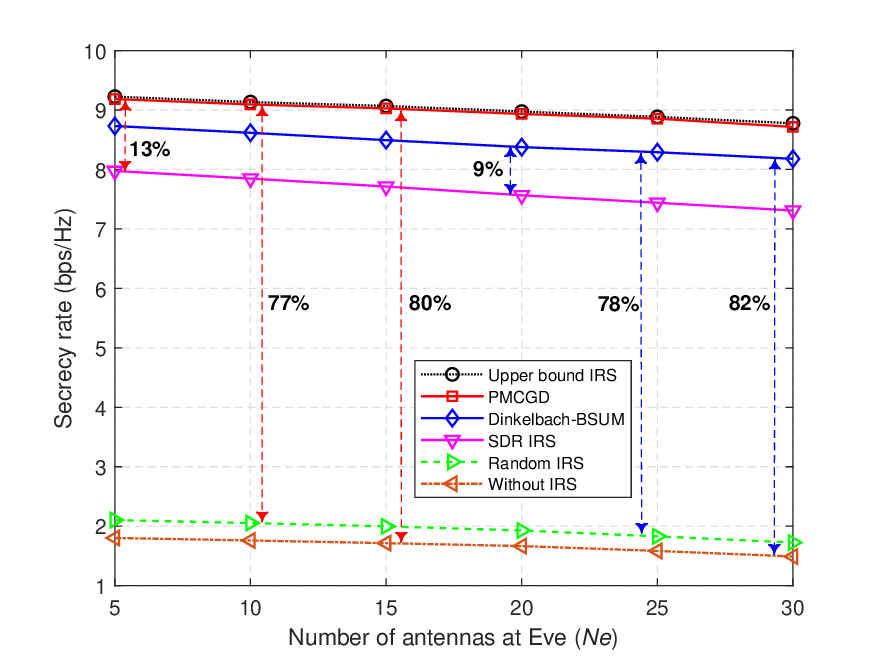}\\
  \caption{ Secrecy rate comparison among different strategies versus the number of antennas at Eve $N_e$.}\label{NumberatEVE}
  \end{center}
\end{figure}

To further examine the proposed algorithms, Figs. \ref{NumberatBOB} and \ref{NumberatEVE} display the average secrecy rate performance with varying numbers of antennas at Bob $N_b$ and Eve $N_e$, respectively. Both figures reveal distinct trends corresponding to the increases in $N_b$ and $N_e$. Specifically, in Fig. \ref{NumberatBOB}, the average secrecy rate for all algorithms increases with the number of Bob's antennas $N_b$. Conversely, in Fig. \ref{NumberatEVE}, the average secrecy rate for all algorithms decreases with the number of Eve's antennas $N_e$. Additionally, the two proposed algorithms outshine the SDR IRS algorithm, the Random IRS algorithm, and the Without IRS algorithm. Notably, for both methods in Fig. \ref{NumberatBOB}, the PMCGD algorithm's secrecy rate is higher by approximately 12\%, 77\%, and 81\% respectively, compared to the SDR IRS, Random IRS, and Without IRS algorithms. Similarly, in Fig. \ref{NumberatEVE}, the PMCGD algorithm maintains higher secrecy rates by roughly 7\%, 76\%, and 79\%, respectively. Furthermore, the Dinkelbach-BSUM algorithm also surpasses the SDR IRS, Random IRS, and Without IRS algorithms in secrecy rate performance, achieving an approximate 13\%, 77\%, and 80\% increase in Fig. \ref{NumberatBOB}, and 9\%, 78\%, and 82\% in Fig. \ref{NumberatEVE}, respectively.

\begin{figure}[htbp]
  \begin{center}
  \includegraphics[width=3.5in]{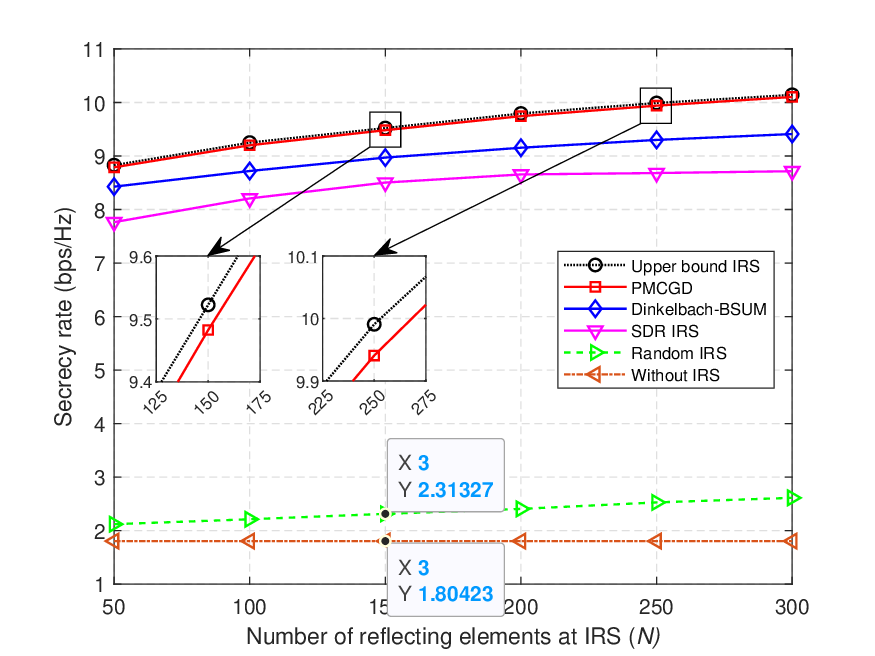}\\
  \caption{ Secrecy rate comparison among different strategies versus the number of IRS reflecting elements $N_i$.}\label{SRIRSN}
  \end{center}
\end{figure}

Fig. \ref{SRIRSN} further demonstrates a performance comparison among different strategies with varying numbers of IRS reflecting elements $N_i$. As observed, all algorithms experience an increase in the secrecy rate with the rise of $N_i$, except for the strategy without IRS. This suggests that the IRS provides the system with more degrees of freedom, thereby enhancing overall performance. When $N_i=150$, the proposed PMCGD and Dinkelbach-BSUM algorithms achieve an average secrecy rate of about 9.5 bps/Hz and 9 bps/Hz, respectively, surpassing the SDR IRS algorithm (8.5 bps/Hz), the Random IRS algorithm (2.3 bps/Hz), and the Without IRS algorithm (1.8 bps/Hz).

The simulation results above highlight the distinct advantages of the proposed algorithms. The Dinkelbach-BSUM algorithm can achieve a higher convergence speed, albeit with a lower average secrecy rate. On the other hand, the PMCGD algorithm can attain an average secrecy rate that is nearly equivalent to the upper bound, but this comes at the cost of increased overall computation time.

\section{Conclusion}
In this paper, we focus on the physical layer security (PLS) of the intelligent reflecting surface (IRS)-assisted large-scale antenna array (LSAA) system. Our goal is to design the constant modulus analog beamforming (CMAB) at the transmitter and the passive beamforming at the IRS so that the multicast secrecy rate is maximized. This secrecy rate maximization problem is generally non-convex due to the highly coupled variables and constant modulus constraints. To solve the problem, two algorithms namely the Dinkelbach-BSUM algorithm and the PMCGD algorithm over the Euclidean and Riemannian spaces are proposed, respectively. Simulation results demonstrate the distinct advantages of the proposed methods.

\begin{appendices} 
\section{proof of theorem \ref{theorem1}}
\label{appendixA}
Before presenting the stationary convergence, we first establish Lemma \ref{proofunique}, which demonstrates the uniqueness of the minimizers for the optimization sub-problems (\ref{usetubw}) and (\ref{Realmintheta}) with respect to the variables \(\mathbf{w}\) and \(\boldsymbol{\theta}\), and Lemma \ref{proofrc}, which addresses the regularity condition of the optimization problem (\ref{NewminDM}). These lemmas will be used later in this section to support the convergence to a stationary point.

\begin{lemma} \label{proofunique}
\textit{(Uniqueness of Minimizers)} The objective $f({\bf w}, {\bm \theta})$ has an unique minimizer for ${\bf w}$ given ${\bm \theta}$, and for ${\bm \theta}$ given ${\bf w}$.
\end{lemma}

\textit{Proof}: \textit{Optimization of ${\bf w}$ given ${\bm \theta}$:} We show that $ {\bf w} = P \text{exp}({j\arg( {\hat {\bf w}} )})$ in (\ref{Closedsw}), where ${\hat {\bf w}}=(\lambda_{\text{max}}({\bf A}){\bf I}-{\bf A}){\bf w}^k$ is the unique minimum for the sub-problem (\ref{usetubw}), by contradiction. 

Let $ {\bf w}^+ = P \text{exp}({j\arg( {\hat {\bf w}} )+{\bm \iota}})$ be another minimum, where at least one entry of ${\bm \iota}$ is non-zero and is not an integer multiple of $2\pi$ (otherwise, ${\bf w}={\bf w}^+$). Then, based on the objective function in (15), we have,
\begin{subequations}
    \begin{align}
  & {g}_{\bf w}({\bf w}) = \sum_{i=1}^M\text{Real}({ w}_i^H{\hat { w_i}}), \label{subfi1} \\
  &  {g}_{\bf w}({\bf w}^+) = \sum_{i=1}^M\text{Real}({w}_i^H{\hat { w_i}})\cos({ \iota_i}).\label{subfi2}
\end{align}
\end{subequations}
Since both ${\bf w}$ and ${\bf w}^+$ minimize ${g}_{\bf w}({\bf w})$, it follows from (\ref{subfi1}) and (\ref{subfi2}) that, ${g}_{\bf w}({\bf w})={g}_{\bf w}({\bf w}^+)$. This is possible only if $\cos({ \iota_i}) ={1},\forall i$, which implies that all the entries of ${\bm \iota}$ are zero or some integer multiples of $2\pi$ going against the assumption. This leads to a contradiction and hence proving the uniqueness. 

\textit{Optimization of ${\bm \theta}$ given ${\bf w}$:} The verification for the optimization of \(\boldsymbol{\theta}\) given \(\mathbf{w}\) follows a similar approach to the proof for the optimization of \(\mathbf{w}\) given \(\boldsymbol{\theta}\). For brevity, we omit the detailed derivation here. $\hfill\blacksquare$

\begin{lemma} \label{proofrc}
\textit{(Regularity Condition)} The function \(f(\mathbf{w}, \boldsymbol{\theta})\) is regular with respect to \(\mathbf{w}\) and \(\boldsymbol{\theta}\).
\end{lemma}

\textit{Proof}: The regularity condition, linear independence constraint qualification (LICQ) implies that the gradients of the active inequality constraints and the gradients of equality constraints are linearly independent at a feasible point (Proposition 3.1.1 in \cite{bertsekas1997nonlinear}). This condition is automatically satisfied at the solution as the constraint sets involving $\bf w$ and $\bm \theta$ satisfy the LICQ condition. For $\bf w$ and $\bm \theta$, the constraints $|w_i|=P$, for all $i \in [M]$ and $|\theta_j|=1$, for all $j \in [N_i]$ are decoupled among the entries $w_i$’s and $\theta_j$’s, respectively. All the constrained optimization variables can be combined in to a single vector as, ${\bf x}=[{\bf w}^T,{\bm \theta}^T]^T$. Now the gradient of constraint $|x_i|^2 - 1=0$ with respect to $\bf x$ is computed as, ${\bf g}_i({\bf x})=2{\bf e}_i$, for all $i \in [M + N]$, where ${\bf e}_i \in {\mathbb R}^{(M+N) \times 1}$ is a vector with $i$-th entry being 1 and rest of the entries are zeros. Clearly all the gradient vectors are linearly independent. This implies that each point of the constraint set of the function \(f(\mathbf{w}, \boldsymbol{\theta})\) is regular. $\hfill\blacksquare$

Now we are ready to prove Theorem \ref{theorem1}. Specifically, we rely on the following theorem from \cite{razaviyayn2013unified}.

\begin{theorem} \label{theoremusesc}
\textit{(Theorem 2(a) in \cite{razaviyayn2013unified})}  
Suppose the surrogate functions in (\ref{usetubw}) and (\ref{Realmintheta}) are quasi-convex with respect to \(\mathbf{w}\) and \(\boldsymbol{\theta}\), respectively, and the properties in (\ref{Assumptiontub}) hold. Furthermore, assume that the subproblems (\ref{usetubw}) and (\ref{Realmintheta}) each have a unique solution for any point satisfying \(|w_i| = P, \forall i\), and \(|\theta_j| = 1, \forall j\). Then, every limit point \(\{\mathbf{w}, \boldsymbol{\theta}\}\) of the iterates generated by the BSUM algorithm is a coordinatewise minimum of (\ref{NewminDM}). Additionally, if \(f(\cdot)\) is regular at \(\{\mathbf{w}, \boldsymbol{\theta}\}\), then \(\{\mathbf{w}, \boldsymbol{\theta}\}\) is a stationary point of (\ref{NewminDM}).
\end{theorem}

Since the uniqueness of minimizers and the regularity condition have been established in Lemma \ref{proofunique} and Lemma \ref{proofrc}, we conclude that every limit point of the sequence generated by the BSUM-based algorithm is a stationary point of the problem, as stated in Theorem \ref{theoremusesc}. This completes the proof. $\hfill\blacksquare$

\section{proof of theorem \ref{theorem2}}
\label{appendixB}
To establish the proof, we first demonstrate that the algorithm achieves a sufficient decrease in each iteration. Specifically, by applying the Armijo line search strategy in (\ref{usecgdm}), we have \cite{boumal2023introduction},
\begin{equation}
\begin{aligned}
  f({\bf w}^k, {\bm \theta}^k) - f({\bf w}^{k+1}, {\bm \theta}^{k+1}) \ge c_{\text{dec}} \| \operatorname{grad}f({\bf w}^k, {\bm \theta}^k) \|_2^2, \label{therem4proof}
\end{aligned}
\end{equation}
where $c_{\text{dec}}  >0$. According to (\ref{therem4proof}), we conclude that the algorithm achieves a sufficient decrease at each iteration.

We can now complete the proof. The proof is based on a standard telescoping sum argument. The desired inequality for all $k=0,1,\dots,K-1$ is obtained as follows,
\begin{subequations}
    \begin{align}
        f({\bf w}^0, {\bm \theta}^0) - &f_{\text{low}}  \ge f({\bf w}^0, {\bm \theta}^0) - f({\bf w}^K, {\bm \theta}^K) \\ & = \sum\limits_{k=0}^{K-1} f({\bf w}^k, {\bm \theta}^k) - f({\bf w}^{k+1}, {\bm \theta}^{k+1}) \\& \ge K c_{\text{dec}} \min _{k=0,1,\dots,K-1}||\operatorname{grad}f({\bf w}_k, {\bf p}_k) \|_2^2
    \end{align}
\end{subequations}
where $f_{\text{low}}=0$ is the lower bound value for the objective function. To get the limit statement, observe that $f({\bf w}^{k+1}, {\bm \theta}^{k+1}) \le f({\bf w}^k, {\bm \theta}^k)$ for all $k$ by (\ref{therem4proof}). Then, taking $K$ to infinity we see that,
\begin{equation}
    f({\bf w}^0, {\bm \theta}^0) -  f_{\text{low}} \ge \sum\limits_{k=0}^{\infty} f({\bf w}^k, {\bm \theta}^k) - f({\bf w}^{k+1}, {\bm \theta}^{k+1}),
\end{equation}
where the right-hand side is a series of nonnegative numbers. The bound implies that the summands converge to zero, thus,
\begin{equation}
\begin{aligned}
    0 &= \lim_{k \to \infty} f({\bf w}^k, {\bm \theta}^k) - f({\bf w}^{k+1}, {\bm \theta}^{k+1}) \\ &\le c_{\text{dec}} \lim_{k \to \infty}\|\operatorname{grad}f({\bf w}^k, {\bm \theta}^k) \|_2^2,
    \end{aligned}
\end{equation}
which confirms that $\|\operatorname{grad}f({\bf w}^k, {\bm \theta}^k) \|_2\to0$. Now, let $\{{\bf w},{\bm \theta}\}$ be a limit point of the sequence of iterates. By definition, there exists a subsequence of iterates
$\{{\bf w}^{(0)},{\bm \theta}^{(0)}\},\{{\bf w}^{(1)},{\bm \theta}^{(1)}\},\{{\bf w}^{(2)},{\bm \theta}^{(2)}\},\dots$ which converges to $\{{\bf w},{\bf p}\}$. Then, since the norm of the gradient of $ f({\bf w}, {\bm \theta})$ is a continuous function, it commutes with the limit and we find,
\begin{equation}
\begin{aligned}
    0 &= \lim_{k \to \infty}\|\operatorname{grad}f({\bf w}^k, {\bm \theta}^k) \|_2^2 = \lim_{k \to \infty}\|\operatorname{grad}f({\bf w}^{(k)}, {\bm \theta}^{(k)}) \|_2^2 \\ & = \|\operatorname{grad}f\left(\lim_{k \to \infty}( {\bf w}^{(k)}, {\bm \theta}^{(k)})\right) \|_2^2 = \|\operatorname{grad}f( {\bf w}, {\bm \theta}) \|_2^2,
\end{aligned}
\end{equation}
showing that all limit points generated by the sequence are stationary points. This completes the proof. $\hfill\blacksquare$

\section{Computation of the Euclidean Gradients}
\label{Calgra}
\textit{Calculation of} $\text{Grad}_{\bf w}f({\bf w},{\bm \theta})$: To obtain the Euclidean gradient related to ${\bf w}$, we reformualte $f({\bf w},{\bm \theta})$ by regarding ${\bm \theta}$ as a constant, i.e.,
\begin{equation}
    f_{\bf w}({\bf w},{\bm \theta}) = \frac{{\bf w}^H{\bf G}_1{\bf w}}{{\bf w}^H{\bf G}_2{\bf w}},
\end{equation}
where,
\begin{subequations}
\begin{align}
{\bf G}_1&=\frac{1}{M}{\bf I}+(\mathbf{H}_{ae} +\mathbf{H}_{ie} {\bf{\Theta }} \mathbf{H}_{ai})^H(\mathbf{H}_{ae} +\mathbf{H}_{ie} {\bf{\Theta }} \mathbf{H}_{ai}),\\
{\bf G}_2&=\frac{1}{M}{\bf I}+(\mathbf{H}_{ab} +\mathbf{H}_{ib} {\bf{\Theta }} \mathbf{H}_{ai})^H(\mathbf{H}_{ab} +\mathbf{H}_{ib} {\bf{\Theta }} \mathbf{H}_{ai}).
\end{align}
\end{subequations}
The quotient rule says,
\begin{equation}
\frac{d}{dx}\left(\frac{u(x)}{v(x)}\right) = \frac{u'(x)v(x) - u(x)v'(x)}{[v(x)]^2}.
\end{equation}
Consequently, we have,
\begin{equation}
\text{Grad}_{\bf w}f({\bf w},{\bm \theta}) = \frac{2{{\bf G}_1}{\bf w}{\bf w}^H{{\bf G}_2}{\bf w}-2{{\bf G}_2}{\bf w}{\bf w}^H{{\bf G}_1}{\bf w}} {({\bf w}^H{{\bf G}_2}{\bf w})^2}.
\label{Gradwnow}
\end{equation}

\textit{Calculation of} $\text{Grad}_{\bm \theta}f({\bf w},{\bm \theta})$: According to the rule $||{\bf w}||_F^2=\text{Tr}({\bf w}^H{\bf w})$, the formula related to ${\bm \theta}$ is given as,
\begin{equation}
    f_{\bm \theta}({\bf w},{\bm \theta}) = \frac{1+ \text{Tr}\{{\bf{\Theta }} {\bf B} {\bf{\Theta }} {\bf C} + {\bf{\Theta }}^H {\bf J}_2^H + {\bf{\Theta }} {\bf J}_2 +{o} \} }{1+\text{Tr}\{{\bf{\Theta }} {\bf E} {\bf{\Theta }} {\bf C} + {\bf{\Theta }}^H {\bf J}_4^H + {\bf{\Theta }} {\bf J}_4 +{p} \}   },
\end{equation}
where,
\begin{subequations}
\begin{align}
{\bf B}& = {\bf H}_{ie}^H{\bf H}_{ie},\\
{\bf C} &= {\bf H}_{ai}{\bf w}{\bf w}^H{\bf H}_{ai}^H,\\
{\bf J}_2&={\bf H}_{ai}{\bf w}{\bf w}^H{\bf H}_{ae}^H{\bf H}_{ie},\\
{ o} &= {\bf w}^H{\bf H}_{ae}^H{\bf H}_{ae}{\bf w},\\
{\bf E} &= {\bf H}_{ib}^H{\bf H}_{ib},\\
{\bf J}_4&={\bf H}_{ai}{\bf w}{\bf w}^H{\bf H}_{ab}^H{\bf H}_{ib},\\
{p} &= {\bf w}^H{\bf H}_{ab}^H{\bf H}_{ab}{\bf w}. 
\end{align}
\end{subequations}
Moreover, according to \cite{zhong2023joint}, we have,
\begin{subequations}
\begin{align}
&1+\text{Tr}\{ {\bf{\Theta }} {\bf B} {\bf{\Theta }} {\bf C} \} = {\bm \theta}^H (\frac{1}{N}{\bf I}+{\bf B}\odot{\bf C}) {\bm \theta} = {\bm \theta}^H {\bf J}_1 {\bm \theta},\\
&1+\text{Tr}\{ {\bf{\Theta }} {\bf E} {\bf{\Theta }} {\bf C} \} = {\bm \theta}^H (\frac{1}{N}{\bf I}+{\bf E}\odot{\bf C}) {\bm \theta} = {\bm \theta}^H {\bf J}_3 {\bm \theta},\\
&\text{Tr}\{ {\bf{\Theta }}^H {\bf J}_2^H \} = {\bm \theta}^H {\bf J}_2^* ,\\ 
&\text{Tr}\{ {\bf{\Theta }} {\bf J}_2 \} =   {\bf J}_2^T {\bm \theta}, \\
&\text{Tr}\{ {\bf{\Theta }}^H {\bf J}_4^H \} = {\bm \theta}^H {\bf J}_4^* ,\\ 
&\text{Tr}\{ {\bf{\Theta }} {\bf J}_4 \} =   {\bf J}_4^T {\bm \theta}.
\end{align}
\end{subequations}
Consequently, $ f_{\bm \theta}({\bf w},{\bm \theta})$ is reformulated as,
\begin{equation}
    f_{\bm \theta}({\bf w},{\bm \theta}) = \frac{{\bm \theta}^H {\bf J}_1 {\bm \theta} + {\bm \theta}^H {\bf J}_2^* +  {\bf J}_2^T {\bm \theta} + o}{{\bm \theta}^H {\bf J}_3 {\bm \theta} + {\bm \theta}^H {\bf J}_4^* +  {\bf J}_4^T {\bm \theta} + p }.
\end{equation}
Similar to the above, following the quotient rule, we have
\begin{equation}
\begin{aligned}
&\text{Grad}_{\bm \theta}f({\bf w},{\bm \theta}) = \\&
\frac{ \left\{\begin{array}{l} 2({{\bf J}_1}{\bm \theta} + {{\bf J}_2}^*)({\bm \theta}^H{{\bf J}_3}{\bm \theta}+{\bm \theta}^H{{\bf J}_4}^*+{{\bf J}_4}^T{\bm \theta}+o) \\ -2({{\bf J}_3}{\bm \theta} + {{\bf J}_4}^*)({\bm \theta}^H{{\bf J}_1}{\bm \theta}+{\bm \theta}^H{{\bf J}_2}^*+{{\bf J}_2}^T{\bm \theta}+p) \end{array}\right\} }{{\bm \theta}^H{{\bf J}_3}{\bm \theta}+{\bm \theta}^H{{\bf J}_4}^*+{{\bf J}_4}^T{\bm \theta}+o}.
\end{aligned}
\label{Gradthetanow}
\end{equation}
This completes the proof. $\hfill\blacksquare$

\end{appendices}

\bibliographystyle{IEEEtran}
\bibliography{Bibliography}

@article{dong2020enhancing,
  title={Enhancing secure MIMO transmission via intelligent reflecting surface},
  author={Dong, Limeng and Wang, Hui-Ming},
  journal={IEEE Transactions on Wireless Communications},
  volume={19},
  number={11},
  pages={7543--7556},
  year={2020},
  publisher={IEEE}
}

@article{li2019constant,
  title={Constant modulus secure beamforming for multicast massive MIMO wiretap channels},
  author={Li, Qiang and Li, Chao and Lin, Jingran},
  journal={IEEE Transactions on Information Forensics and Security},
  volume={15},
  pages={264--275},
  year={2019},
  publisher={IEEE}
}

@article{yang2023secure,
  title={Secure hybrid beamforming for IRS-assisted millimeter wave systems},
  author={Yang, Long and Wang, Jiangtao and Xue, Xuan and Shi, Jia and Wang, Yongchao},
  journal={IEEE Transactions on Wireless Communications},
  year={2023},
  publisher={IEEE}
}

@article{cui2019secure,
  title={Secure wireless communication via intelligent reflecting surface},
  author={Cui, Miao and Zhang, Guangchi and Zhang, Rui},
  journal={IEEE Wireless Communications Letters},
  volume={8},
  number={5},
  pages={1410--1414},
  year={2019},
  publisher={IEEE}
}

@ARTICLE{5447068,
  author={Luo, Zhi-quan and Ma, Wing-kin and So, Anthony Man-cho and Ye, Yinyu and Zhang, Shuzhong},
  journal={IEEE Signal Processing Magazine}, 
  title={Semidefinite Relaxation of Quadratic Optimization Problems}, 
  year={2010},
  volume={27},
  number={3},
  pages={20-34},
  doi={10.1109/MSP.2010.936019}}

@article{wu2019towards,
  title={Towards smart and reconfigurable environment: Intelligent reflecting surface aided wireless network},
  author={Wu, Qingqing and Zhang, Rui},
  journal={IEEE communications magazine},
  volume={58},
  number={1},
  pages={106--112},
  year={2019},
  publisher={IEEE}
}

@article{cheng2023ris,
  title={RIS-Assisted Secure Communications: Low-Complexity Beamforming Design},
  author={Cheng, Zhenqiao and Li, Nanxi and Zhu, Jianchi and She, Xiaoming and Ouyang, Chongjun and Chen, Peng},
  journal={IEEE Wireless Communications Letters},
  year={2023},
  publisher={IEEE}
}

@article{liu2021reconfigurable,
  title={Reconfigurable intelligent surfaces: Principles and opportunities},
  author={Liu, Yuanwei and Liu, Xiao and Mu, Xidong and Hou, Tianwei and Xu, Jiaqi and Di Renzo, Marco and Al-Dhahir, Naofal},
  journal={IEEE communications surveys \& tutorials},
  volume={23},
  number={3},
  pages={1546--1577},
  year={2021},
  publisher={IEEE}
}

@article{khisti2010secure,
  title={Secure transmission with multiple antennas—Part II: The MIMOME wiretap channel},
  author={Khisti, Ashish and Wornell, Gregory W},
  journal={IEEE Transactions on Information Theory},
  volume={56},
  number={11},
  pages={5515--5532},
  year={2010},
  publisher={IEEE}
}

@article{asaad2022secure,
  title={Secure active and passive beamforming in IRS-aided MIMO systems},
  author={Asaad, Saba and Wu, Yifei and Bereyhi, Ali and Mueller, Ralf R and Schaefer, Rafael F and Poor, H Vincent},
  journal={IEEE Transactions on Information Forensics and Security},
  volume={17},
  pages={1300--1315},
  year={2022},
  publisher={IEEE}
}

@article{zhong2023joint,
  title={Joint waveform and beamforming design for RIS-aided ISAC systems},
  author={Zhong, Kai and Hu, Jinfeng and Pan, Cunhua and Deng, Minglong and Fang, Jun},
  journal={IEEE Signal Processing Letters},
  volume={30},
  pages={165--169},
  year={2023},
  publisher={IEEE}
}

@ARTICLE{6772207,
  author={Wyner, A. D.},
  journal={The Bell System Technical Journal}, 
  title={The wire-tap channel}, 
  year={1975},
  volume={54},
  number={8},
  pages={1355-1387},
  doi={10.1002/j.1538-7305.1975.tb02040.x}}

@article{zhu2014secure,
  title={Secure transmission in multicell massive MIMO systems},
  author={Zhu, Jun and Schober, Robert and Bhargava, Vijay K},
  journal={IEEE Transactions on Wireless Communications},
  volume={13},
  number={9},
  pages={4766--4781},
  year={2014},
  publisher={IEEE}
}

@article{bjornson2020scalable,
  title={Scalable cell-free massive MIMO systems},
  author={Bj{\"o}rnson, Emil and Sanguinetti, Luca},
  journal={IEEE Transactions on Communications},
  volume={68},
  number={7},
  pages={4247--4261},
  year={2020},
  publisher={IEEE}
}

@article{chen2020structured,
  title={Structured massive access for scalable cell-free massive MIMO systems},
  author={Chen, Shuaifei and Zhang, Jiayi and Bj{\"o}rnson, Emil and Zhang, Jing and Ai, Bo},
  journal={IEEE Journal on Selected Areas in Communications},
  volume={39},
  number={4},
  pages={1086--1100},
  year={2020},
  publisher={IEEE}
}

@article{tang2022dilated,
  title={Dilated convolution based CSI feedback compression for massive MIMO systems},
  author={Tang, Shunpu and Xia, Junjuan and Fan, Lisheng and Lei, Xianfu and Xu, Wei and Nallanathan, Arumugam},
  journal={IEEE Transactions on Vehicular Technology},
  volume={71},
  number={10},
  pages={11216--11221},
  year={2022},
  publisher={IEEE}
}

@article{zhu2015linear,
  title={Linear precoding of data and artificial noise in secure massive MIMO systems},
  author={Zhu, Jun and Schober, Robert and Bhargava, Vijay K},
  journal={IEEE Transactions on Wireless Communications},
  volume={15},
  number={3},
  pages={2245--2261},
  year={2015},
  publisher={IEEE}
}

@article{wu2016secure,
  title={Secure massive MIMO transmission with an active eavesdropper},
  author={Wu, Yongpeng and Schober, Robert and Ng, Derrick Wing Kwan and Xiao, Chengshan and Caire, Giuseppe},
  journal={IEEE Transactions on Information Theory},
  volume={62},
  number={7},
  pages={3880--3900},
  year={2016},
  publisher={IEEE}
}

@article{kapetanovic2015physical,
  title={Physical layer security for massive MIMO: An overview on passive eavesdropping and active attacks},
  author={Kapetanovic, Dzevdan and Zheng, Gan and Rusek, Fredrik},
  journal={IEEE Communications Magazine},
  volume={53},
  number={6},
  pages={21--27},
  year={2015},
  publisher={IEEE}
}

@inproceedings{xu2019resource,
  title={Resource allocation for secure IRS-assisted multiuser MISO systems},
  author={Xu, Dongfang and Yu, Xianghao and Sun, Yan and Ng, Derrick Wing Kwan and Schober, Robert},
  booktitle={2019 IEEE Globecom Workshops (GC Wkshps)},
  pages={1--6},
  year={2019},
  organization={IEEE}
}

@article{zappone2015energy,
  title={Energy efficiency in wireless networks via fractional programming theory},
  author={Zappone, Alessio and Jorswieck, Eduard and others},
  journal={Foundations and Trends{\textregistered} in Communications and Information Theory},
  volume={11},
  number={3-4},
  pages={185--396},
  year={2015},
  publisher={Now Publishers, Inc.}
}

@article{razaviyayn2013unified,
  title={A unified convergence analysis of block successive minimization methods for nonsmooth optimization},
  author={Razaviyayn, Meisam and Hong, Mingyi and Luo, Zhi-Quan},
  journal={SIAM Journal on Optimization},
  volume={23},
  number={2},
  pages={1126--1153},
  year={2013},
  publisher={SIAM}
}

@ARTICLE{5605343,
  author={Khisti, Ashish and Wornell, Gregory W.},
  journal={IEEE Transactions on Information Theory}, 
  title={Secure Transmission With Multiple Antennas—Part II: The MIMOME Wiretap Channel}, 
  year={2010},
  volume={56},
  number={11},
  pages={5515-5532},
  doi={10.1109/TIT.2010.2068852}}

@article{song2015optimization,
  title={Optimization methods for designing sequences with low autocorrelation sidelobes},
  author={Song, Junxiao and Babu, Prabhu and Palomar, Daniel P},
  journal={IEEE Transactions on Signal Processing},
  volume={63},
  number={15},
  pages={3998--4009},
  year={2015},
  publisher={IEEE}
}

@article{hunter2004tutorial,
  title={A tutorial on MM algorithms},
  author={Hunter, David R and Lange, Kenneth},
  journal={The American Statistician},
  volume={58},
  number={1},
  pages={30--37},
  year={2004},
  publisher={Taylor \& Francis}
}

@book{lange2016mm,
  title={MM optimization algorithms},
  author={Lange, Kenneth},
  year={2016},
  publisher={SIAM}
}

@book{boumal2023introduction,
  title={An introduction to optimization on smooth manifolds},
  author={Boumal, Nicolas},
  year={2023},
  publisher={Cambridge University Press}
}

@article{yu2016alternating,
  title={Alternating minimization algorithms for hybrid precoding in millimeter wave MIMO systems},
  author={Yu, Xianghao and Shen, Juei-Chin and Zhang, Jun and Letaief, Khaled B},
  journal={IEEE Journal of Selected Topics in Signal Processing},
  volume={10},
  number={3},
  pages={485--500},
  year={2016},
  publisher={IEEE}
}

@article{hu2022constant,
  title={Constant modulus waveform design for MIMO radar via manifold optimization},
  author={Hu, Jinfeng and Zhang, Weijian and Zhu, Haoming and Zhong, Kai and Xiong, Weijie and Wei, Zhiyong and Li, Yuzhi},
  journal={Signal Processing},
  volume={190},
  pages={108322},
  year={2022},
  publisher={Elsevier}
}

@ARTICLE{202312,
  author={Zhong, Kai and Hu, Jinfeng and et al.},
  journal={IEEE Transactions on Aerospace and Electronic Systems}, 
  title={MIMO Radar Unimodular Waveform Design With Learned Complex Circle Manifold Network}, 
  year={2023},
  volume={},
  number={},
  pages={},
  doi={10.1109/TAES.2023.3344391}}

@article{babaie2015hybridization,
  title={A hybridization of the Polak-Ribi{\`e}re-Polyak and Fletcher-Reeves conjugate gradient methods},
  author={Babaie-Kafaki, Saman and Ghanbari, Reza},
  journal={Numerical Algorithms},
  volume={68},
  pages={481--495},
  year={2015},
  publisher={Springer}
}

@article{hong2020artificial,
  title={Artificial-noise-aided secure MIMO wireless communications via intelligent reflecting surface},
  author={Hong, Sheng and Pan, Cunhua and Ren, Hong and Wang, Kezhi and Nallanathan, Arumugam},
  journal={IEEE Transactions on Communications},
  volume={68},
  number={12},
  pages={7851--7866},
  year={2020},
  publisher={IEEE}
}

@article{yu2020robust,
  title={Robust and secure wireless communications via intelligent reflecting surfaces},
  author={Yu, Xianghao and Xu, Dongfang and Sun, Ying and Ng, Derrick Wing Kwan and Schober, Robert},
  journal={IEEE Journal on Selected Areas in Communications},
  volume={38},
  number={11},
  pages={2637--2652},
  year={2020},
  publisher={IEEE}
}

@article{wang2022intelligent,
  title={Intelligent reflecting surface aided secure transmission with colluding eavesdroppers},
  author={Wang, Yang and Shi, Weiping and Huang, Mengxing and Shu, Feng and Wang, Jiangzhou},
  journal={IEEE Transactions on Vehicular Technology},
  volume={71},
  number={9},
  pages={10155--10160},
  year={2022},
  publisher={IEEE}
}

@inproceedings{mukherjee2012detecting,
  title={Detecting passive eavesdroppers in the MIMO wiretap channel},
  author={Mukherjee, Amitav and Swindlehurst, A Lee},
  booktitle={2012 IEEE International Conference on Acoustics, Speech and Signal Processing (ICASSP)},
  pages={2809--2812},
  year={2012},
  organization={IEEE}
}

@article{arora2021efficient,
  title={Efficient algorithms for constant-modulus analog beamforming},
  author={Arora, Aakash and Tsinos, Christos G and Shankar, MR Bhavani and Chatzinotas, Symeon and Ottersten, Bj{\"o}rn},
  journal={IEEE Transactions on Signal Processing},
  volume={70},
  pages={756--771},
  year={2021},
  publisher={IEEE}
}

@article{pradhan2020hybrid,
  title={Hybrid precoding design for reconfigurable intelligent surface aided mmWave communication systems},
  author={Pradhan, Chandan and Li, Ang and Song, Lingyang and Vucetic, Branka and Li, Yonghui},
  journal={IEEE Wireless Communications Letters},
  volume={9},
  number={7},
  pages={1041--1045},
  year={2020},
  publisher={IEEE}
}

@article{xiu2021reconfigurable,
  title={Reconfigurable intelligent surfaces aided mmWave NOMA: Joint power allocation, phase shifts, and hybrid beamforming optimization},
  author={Xiu, Yue and Zhao, Jun and Sun, Wei and Di Renzo, Marco and Gui, Guan and Zhang, Zhongpei and Wei, Ning},
  journal={IEEE Transactions on Wireless Communications},
  volume={20},
  number={12},
  pages={8393--8409},
  year={2021},
  publisher={IEEE}
}

@article{di2020hybrid,
  title={Hybrid beamforming for reconfigurable intelligent surface based multi-user communications: Achievable rates with limited discrete phase shifts},
  author={Di, Boya and Zhang, Hongliang and Song, Lingyang and Li, Yonghui and Han, Zhu and Poor, H Vincent},
  journal={IEEE Journal on Selected Areas in Communications},
  volume={38},
  number={8},
  pages={1809--1822},
  year={2020},
  publisher={IEEE}
}

@article{li2022joint,
  title={Joint design of hybrid beamforming and reflection coefficients in RIS-aided mmWave MIMO systems},
  author={Li, Renwang and Guo, Bei and Tao, Meixia and Liu, Ya-Feng and Yu, Wei},
  journal={IEEE Transactions on Communications},
  volume={70},
  number={4},
  pages={2404--2416},
  year={2022},
  publisher={IEEE}
}

@article{zhou2021stochastic,
  title={Stochastic learning-based robust beamforming design for RIS-aided millimeter-wave systems in the presence of random blockages},
  author={Zhou, Gui and Pan, Cunhua and Ren, Hong and Wang, Kezhi and Elkashlan, Maged and Di Renzo, Marco},
  journal={IEEE Transactions on Vehicular Technology},
  volume={70},
  number={1},
  pages={1057--1061},
  year={2021},
  publisher={IEEE}
}

@article{chen2023irs,
  title={IRS-aided joint spatial division and multiplexing for mmWave multiuser MISO systems},
  author={Chen, Zijian and Zhao, Ming-Min and Li, Min and Lei, Ming and Zhao, Min-Jian},
  journal={IEEE Transactions on Wireless Communications},
  volume={22},
  number={11},
  pages={7789--7804},
  year={2023},
  publisher={IEEE}
}

@article{leung1978gaussian,
  title={The Gaussian wire-tap channel},
  author={Leung-Yan-Cheong, S and Hellman, Ma},
  journal={IEEE transactions on information theory},
  volume={24},
  number={4},
  pages={451--456},
  year={1978},
  publisher={IEEE}
}

@article{sohrabi2016hybrid,
  title={Hybrid digital and analog beamforming design for large-scale antenna arrays},
  author={Sohrabi, Foad and Yu, Wei},
  journal={IEEE Journal of Selected Topics in Signal Processing},
  volume={10},
  number={3},
  pages={501--513},
  year={2016},
  publisher={IEEE}
}

@article{ahmed2018survey,
  title={A survey on hybrid beamforming techniques in 5G: Architecture and system model perspectives},
  author={Ahmed, Irfan and Khammari, Hedi and Shahid, Adnan and Musa, Ahmed and Kim, Kwang Soon and De Poorter, Eli and Moerman, Ingrid},
  journal={IEEE Communications Surveys \& Tutorials},
  volume={20},
  number={4},
  pages={3060--3097},
  year={2018},
  publisher={IEEE}
}

@article{molisch2017hybrid,
  title={Hybrid beamforming for massive MIMO: A survey},
  author={Molisch, Andreas F and Ratnam, Vishnu V and Han, Shengqian and Li, Zheda and Nguyen, Sinh Le Hong and Li, Linsheng and Haneda, Katsuyuki},
  journal={IEEE Communications magazine},
  volume={55},
  number={9},
  pages={134--141},
  year={2017},
  publisher={IEEE}
}

@article{lin2019hybrid,
  title={Hybrid beamforming for millimeter wave systems using the MMSE criterion},
  author={Lin, Tian and Cong, Jiaqi and Zhu, Yu and Zhang, Jun and Letaief, Khaled Ben},
  journal={IEEE Transactions on Communications},
  volume={67},
  number={5},
  pages={3693--3708},
  year={2019},
  publisher={IEEE}
}

@article{huang2021navigation,
  title={Navigation of a UAV team for collaborative eavesdropping on multiple ground transmitters},
  author={Huang, Hailong and Savkin, Andrey V and Ni, Wei},
  journal={IEEE Transactions on Vehicular Technology},
  volume={70},
  number={10},
  pages={10450--10460},
  year={2021},
  publisher={IEEE}
}

@article{yuan2019secrecy,
  title={Secrecy performance of terrestrial radio links under collaborative aerial eavesdropping},
  author={Yuan, Xin and Feng, Zhiyong and Ni, Wei and Liu, Ren Ping and Zhang, J Andrew and Xu, Wenjun},
  journal={IEEE Transactions on Information Forensics and Security},
  volume={15},
  pages={604--619},
  year={2019},
  publisher={IEEE}
}

@article{zhang2024physical,
  title={Physical layer security in near-field communications},
  author={Zhang, Zheng and Liu, Yuanwei and Wang, Zhaolin and Mu, Xidong and Chen, Jian},
  journal={IEEE Transactions on Vehicular Technology},
  year={2024},
  publisher={IEEE}
}

@article{chen2024physical,
  title={Physical Layer Security for Near-Field Communications via Directional Modulation},
  author={Chen, Jiangong and Xiao, Yue and Liu, Kanglai and Zhong, Yuan and Lei, Xia and Xiao, Ming},
  journal={IEEE Transactions on Vehicular Technology},
  year={2024},
  publisher={IEEE}
}

@article{dai2022unmanned,
  title={Unmanned-aerial-vehicle-assisted wireless networks: Advancements, challenges, and solutions},
  author={Dai, Minghui and Huang, Ning and Wu, Yuan and Gao, Jie and Su, Zhou},
  journal={IEEE Internet of Things Journal},
  volume={10},
  number={5},
  pages={4117--4147},
  year={2022},
  publisher={IEEE}
}

@article{adil2023uav,
  title={UAV-assisted IoT applications, cybersecurity threats, AI-enabled solutions, open challenges with future research directions},
  author={Adil, Muhammad and Song, Houbing and Mastorakis, Spyridon and Abulkasim, Hussein and Farouk, Ahmed and Jin, Zhanpeng},
  journal={IEEE Transactions on Intelligent Vehicles},
  volume={9},
  number={4},
  pages={4583--4605},
  year={2023},
  publisher={IEEE}
}

@article{bertsekas1997nonlinear,
  title={Nonlinear programming},
  author={Bertsekas, Dimitri P},
  journal={Journal of the Operational Research Society},
  volume={48},
  number={3},
  pages={334--334},
  year={1997},
  publisher={Taylor \& Francis}
}

@article{liu2024survey,
  title={A survey of recent advances in optimization methods for wireless communications},
  author={Liu, Ya-Feng and Chang, Tsung-Hui and Hong, Mingyi and Wu, Zheyu and So, Anthony Man-Cho and Jorswieck, Eduard A and Yu, Wei},
  journal={IEEE Journal on Selected Areas in Communications},
  year={2024},
  publisher={IEEE}
}

\vfill

\end{document}